\documentclass[10pt,aps,prx,twocolumn,showpacs,preprintnumbers,amsmath,amssymb,superscriptaddress]{revtex4-2}

\usepackage{lineno}
\usepackage{lipsum}

\usepackage{graphicx}
\usepackage{dcolumn}
\usepackage{bm}

\usepackage{hyperref}
\hypersetup{
	colorlinks,
	citecolor=red,
	filecolor=black,
	linkcolor=blue,
	urlcolor=blue
}
\usepackage{natbib}
\usepackage{soul}

\usepackage{dsfont}
\usepackage{tgtermes}

\newcommand{\beq}{\begin{equation}}
\newcommand{\eeq}{\end{equation}}
\newcommand{\btcb}[2]{\begin{tcolorbox}[arc=#1mm,boxsep=#2mm]}
\newcommand{\etcb}{\end{tcolorbox}}
\newcommand{\bd}{\textbf}

\newcommand{\bds}{\boldsymbol}
\newcommand{\derd}{\, \mathrm d}

\newcommand{\ham}{\mathcal{H}}
\newcommand{\hham}{\hat{\mathcal{H}}}

\newcommand{\la}{\langle}

\newcommand{\ra}{\rangle}

\newcommand{\tit}{\textit}

\newcommand{\sgn}{\text{sgn}}

\newcommand{\tr}{\text{Tr}}

\usepackage{graphicx}
\usepackage{dcolumn}
\usepackage{bm}

\begin{document}



\title{Magnetic Field Walls in Flat-band Superconductors}
\author{Guodong Jiang}
\email{guodong.jiang@aalto.fi}
\affiliation{Department of Applied Physics, Aalto University School of Science, FI-00076 Aalto, Finland}
\author{Aaron Dunbrack}
\affiliation{Department of Physics and Nanoscience Center, University of Jyv{\"a}skyl{\"a}, FI-40014 University of Jyv\"askyl\"a, Finland}
\author{Tero T. Heikkil{\"a}}
\affiliation{Department of Physics and Nanoscience Center, University of Jyv{\"a}skyl{\"a}, FI-40014 University of Jyv\"askyl\"a, Finland}
\author{P{\"a}ivi T{\"o}rm{\"a}}
\email{paivi.torma@aalto.fi}
\affiliation{Department of Applied Physics, Aalto University School of Science, FI-00076 Aalto, Finland}

\begin{abstract}
Superconductors of different types have distinct magnetic properties; for example, they can form Abrikosov vortices or alternating normal-superconducting domains. We predict that, in flat bands, a superconducting phase exhibiting walls of magnetic flux is stable in an applied magnetic field. This phase relies on the lack of a single-particle energy penalty for forming condensates of any momentum in flat bands and, consequently, their superconducting free energy being a negative and periodic function of the vector potential at low temperatures. Using a minimal lattice-periodic model of free energy, we study two types of soliton modes of the wall phase: the kink and breather solitons. They determine the lower critical field and the high-field behavior of the wall phase, respectively. The competition between the walls and vortices in flat bands is also discussed. Our results suggest that flat bands help sustain superconductivity in the presence of large magnetic fields.
\end{abstract}
\maketitle
\section{Introduction}
Superconductors can be classified into different types according to their magnetic properties, and show distinct structures of magnetic flux~\cite{campbell1972,blatter1994,abrikosov2004nobel,babaev2005semi,moshchalkov2009type,agterberg2014microscopic}. These classifications are based on the ratio between the penetration depth and the coherence length, which are derived from phenomenological theories typically assuming dispersive bands. Similar understanding does not yet exist for flat bands, which are of interest due to the superconductivity observed in twisted two-dimensional (2D) crystals~\cite{bistritzer2011moire,cao2018unconventional,cao2018correlated,andrei2021marvels,balents2020superconductivity,kennes2021,torma2022tbgreview} and the role of quantum geometry~\cite{provost1980riemannian,resta2011insulating} uncovered in superfluidity~\cite{peotta2015superfluidity,liang2017band,yu2025quantum}.

In dispersive bands, a condensate of finite momentum costs kinetic energy. On the contrary, in flat bands, the lack of a Fermi surface and absence of such a cost allows forming condensates of any momentum. Recent studies on the density-wave orders in multi-orbital flat bands~\cite{jiang2023pdw,chen2023pair,han2024qgn,sun2025flat,wang2025density,lamponen2025super,zhang2026identifying} have revealed that in some lattice models superconductivity is energetically favoured for condensate momenta
in the entire Cooper-pair Brillouin zone~\cite{jiang2023pdw,sun2025flat,dunbrack2026trsbreak}. Furthermore, the pairing susceptibility at high temperatures depends on the condensate momentum only through the Bloch wavefunction form factors~\cite{han2024qgn,sun2025flat,zhang2026identifying}. These features reflect the flexibility of flat bands in supporting condensates of any momentum, in contrast to dispersive bands. In the presence of a magnetic field, the magnetic vector potential $\bd{A}$ couples with the crystal momentum in the same way as the condensate momentum, $\bd{Q}$, by replacing $\bd{Q}\rightarrow(2e/\hbar)\bd{A}$, with $e$ the electron charge and $\hbar$ the reduced Planck constant. Therefore, the above properties imply that flat-band superconductors, compared to dispersive band ones, may respond in a fundamentally different way to an external magnetic field.

\begin{figure}[t!]
\centering	
\includegraphics[width=0.48\textwidth]{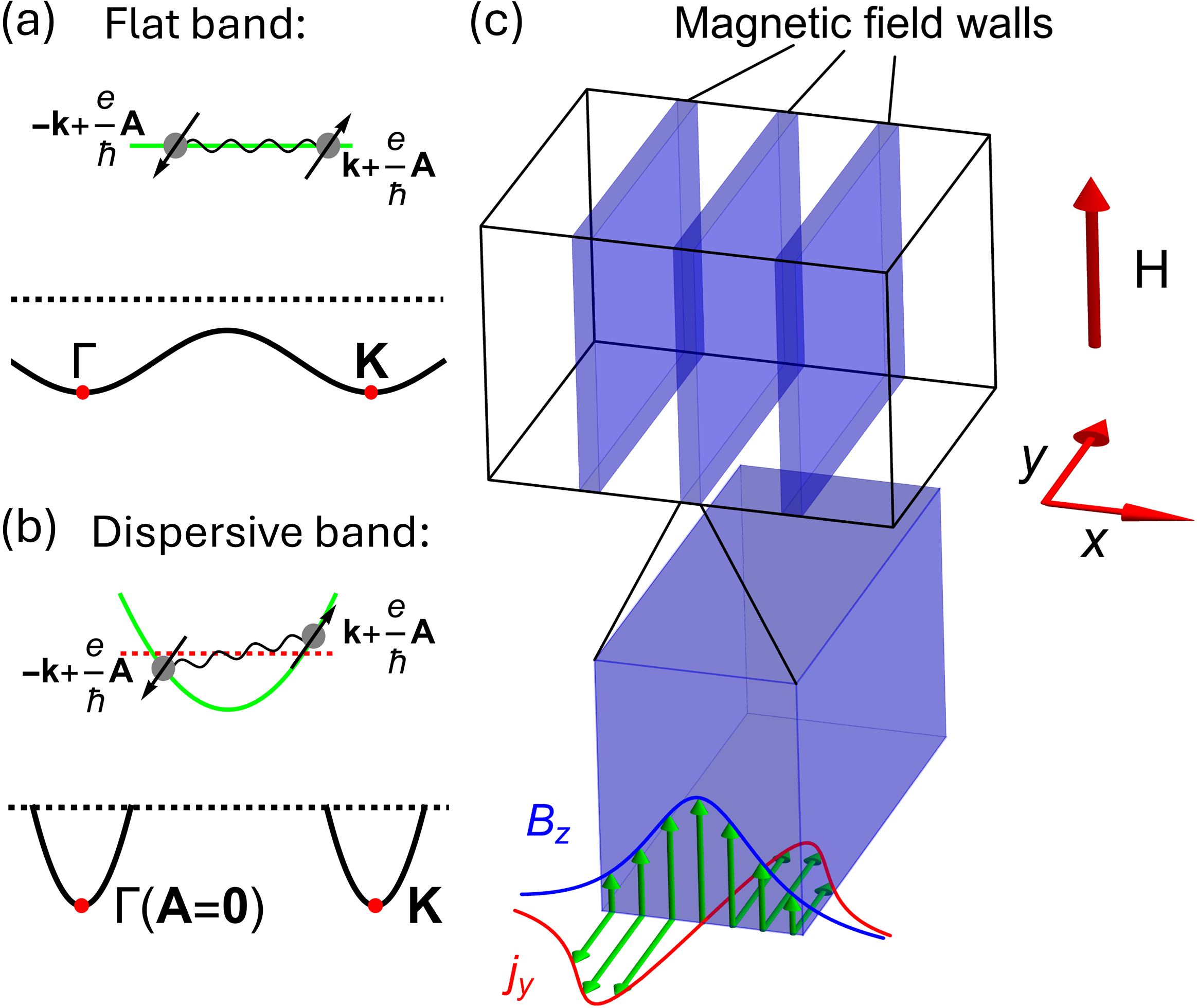}
\caption{(a)-(b) Schematic of pairing with a nonzero vector potential $\bd{A}$, and the free energy density of superconductors, $f_s(\bd{A})$, along a high-symmetry direction at low temperatures, in (a) a flat and (b) a dispersive band. Upper panels: the bands are shown in green, with the Fermi surface of the dispersive case indicated by the red dashed line. The wiggly lines represent the attractive interaction between spin-$\uparrow$ and $\downarrow$ electrons (spheres). Lower panels: $f_s(\bd{A})$ has periodicity $\hbar\bd{K}/e$, where $\bd{K}$ denotes a time-reversal invariant momentum; the black dashed line indicates the zero-energy point, $f_s=0$, above which the superconducting state is unstable. (c) Schematic of a flat-band superconducting bulk material in an applied magnetic field $\bd{H}=H\hat{\bd{z}}$. The magnetic field walls (concentrated in the blue regions) arise from the global negativity of $f_s(\bd{A})$ in (a). The magnified picture shows the shape of a single magnetic field wall with the vector fields of the $B$ field $B_z\hat{\bd{z}}$ and the current density $j_y\hat{\bd{y}}$ represented by green arrows.}
\label{fig:bulk}
\end{figure}

In this Article, we predict a superconducting phase exhibiting walls of magnetic flux in the bulk of the superconductor. The wall phase emerges when the free energy density $f_s(\bd{A})$ of the superconductor remains negative for all vector potentials $\bd{A}$ along a high-symmetry direction of the crystal. This typically happens in multi-orbital flat bands because the wavefunction form factor, rather than the kinetic energy as in dispersive bands, determines the depairing of the order parameter; see Fig.~\ref{fig:bulk}(a)-(b) for an illustration of the pairing and free energy density in the presence of a magnetic field for flat and dispersive bands. The global negativity of the free energy density leads to two types of soliton solutions, kink and breather solitons, to the Maxwell equations, corresponding to periodically arranged ``walls" of magnetic flux inside the superconducting bulk, see Fig.~\ref{fig:bulk}(c). At lower magnetic fields, the walls are separated, with the wall thickness given by the penetration depth; at higher ones, they merge so that the field penetrates the whole bulk, and diamagnetism gradually vanishes. The comparison between the kink state and the uniform superconducting state determines the lower critical field, while the comparison between the breather state and the normal state, without considering other microscopic ingredients, shows an absence of the upper critical field for the wall phase. The walls can form textures, e.g., grids. The wall solution is favored over vortices with a large core size due to the high energy cost of the vortex core in a flat-band superconductor. Our findings introduce a mechanism for the coexistence of superconductivity and magnetism that originates from band quantum geometry rather than the spin symmetry of pairing or the creation of normal state regions, e.g., vortex cores.

\section{Magnetic field dependence of the free energy in a flat band}
We now briefly discuss the periodicity of the free energy density and then analyse the superconducting state in the presence of a magnetic field, using a simple but generic \textit{cosine model} that captures the negativity and periodicity of the free energy density. The free energy density $f_s(\bd{A})$ has the periodicity of the Cooper-pair Brillouin zone defined by a time-reversal invariant momentum (TRIM) $\bd{K}$, i.e., when $\bd{K}$ and $-\bd{K}$ differ by a reciprocal lattice vector. This periodicity arises because the two ways of singlet pairing, $(\bd{k}+\frac{e}{\hbar}\bd{A}\uparrow,-\bd{k}+\frac{e}{\hbar}\bd{A}\downarrow)$ and $(\bd{k}+\frac{e}{\hbar}\bd{A}+\bd{K}\uparrow,-\bd{k}+\frac{e}{\hbar}\bd{A}+\bd{K}\downarrow)$ are equivalent. Here, $\bd{A}$ is \tit{gauge-invariant} since we remove the phase gradient from the superconducting order parameter. The function $f_s(\bd{A})$ of a model Hamiltonian can be calculated from the grand potential formalism in a locally self-consistent manner (see Appendix~\ref{app:fsdensity}).

We consider a superconducting bulk material with geometry depicted in Fig.~\ref{fig:bulk}(c). The magnetic field $\bd{H}$ is applied along a long axis ($z$ axis), and we disregard the edge effects distorting the field direction. The superconductivity is assumed to occur in a band which is flat in the $k_x-k_y$ plane. The vector potential $\bd{A}$ is restricted in the $xy$ plane since its $z$-component vanishes. The current density is given by (see Appendix~\ref{app:gauge} for details)
\beq\label{eq:jdef}
\bd{j}(\bd{r})=-\nabla_\bd{A}f_s(\bd{A}(\bd{r})).
\eeq
We then solve the Maxwell equations
\beq\label{eq:maxwell}
\nabla\times\bd{A}(\bd{r})=\bd{B}(\bd{r}),\,\,\,\nabla\times\bd{B}(\bd{r})=\mu_0\bd{j}(\bd{r}),
\eeq
($\mu_0$ is the vacuum permeability) together with Eq.~\eqref{eq:jdef} for the fields $\bd{A}(\bd{r})$, $\bd{B}(\bd{r})$.

As an ansatz, we seek a continuum solution of the form $\bd{A}\parallel\hat{\bd{y}}$ and with translational symmetry along the $y$ axis. We also assume that the $y$ axis is parallel to a high-symmetry direction of the crystal, along which $\bd{j}\parallel\bd{A}$. These assumptions imply $\bd{A}(x)=A_y(x)\hat{\bd{y}}$, $\bd{B}(x)=B_z(x)\hat{\bd{z}}$ and $\bd{j}(x)=j_y(x)\hat{\bd{y}}$. Then Eq.~\eqref{eq:maxwell} leads to a second-order differential equation of $A_y$,
\beq\label{eq:maxwell2}
A_y''(x)-\mu_0\partial_{A_y}f_s(A_y(x))=0,
\eeq
where $f_s(A_y)$ stands for $f_s(A_x=0,A_y)$. In the London theory of superconductors, one uses
\beq\label{eq:london}
f_s(A_y)\approx f_s(0)+\frac{1}{2}D_sA_y^2,
\eeq
with $D_s$ the superfluid weight (stiffness), to get the exponential solution, $A_y(x)\propto \exp\{\pm\sqrt{\mu_0D_s}x\}$. For flat bands, instead, we consider the following \textit{cosine model},
\beq\label{eq:cosine}
f_s(A_y)=-\delta_1-\delta_2\cos\left(\frac{2ea}{\hbar}A_y\right),
\eeq
where the two parameters $\delta_1,\delta_2$ ($\delta_1>\delta_2>0$) represent the average and oscillation amplitude of the condensation energy density along the high-symmetry line, respectively. We assume that the neighboring TRIMs in the $k_x-k_y$ plane are separated by $\pi/a$, with $a$ the lattice constant in the $xy$ plane, e.g., considering the tetragonal lattice case. This model captures a few salient features. First, the $\Gamma$-point $\bd{A}=0$ and its replica at TRIMs are the global minima of free energy, where the condensation energy density is $\delta_1+\delta_2$. Second, $f_s(A_y)$ is negative, and its oscillatory behavior is simplified as a cosine function. Third, near the $\Gamma$-point, $f_s(A_y)$ reduces to the London relation, Eq.~\eqref{eq:london}, with $D_s=4(ea/\hbar)^2\delta_2$.  

\begin{figure}[t!]
\centering
\includegraphics[width=0.3\textwidth]{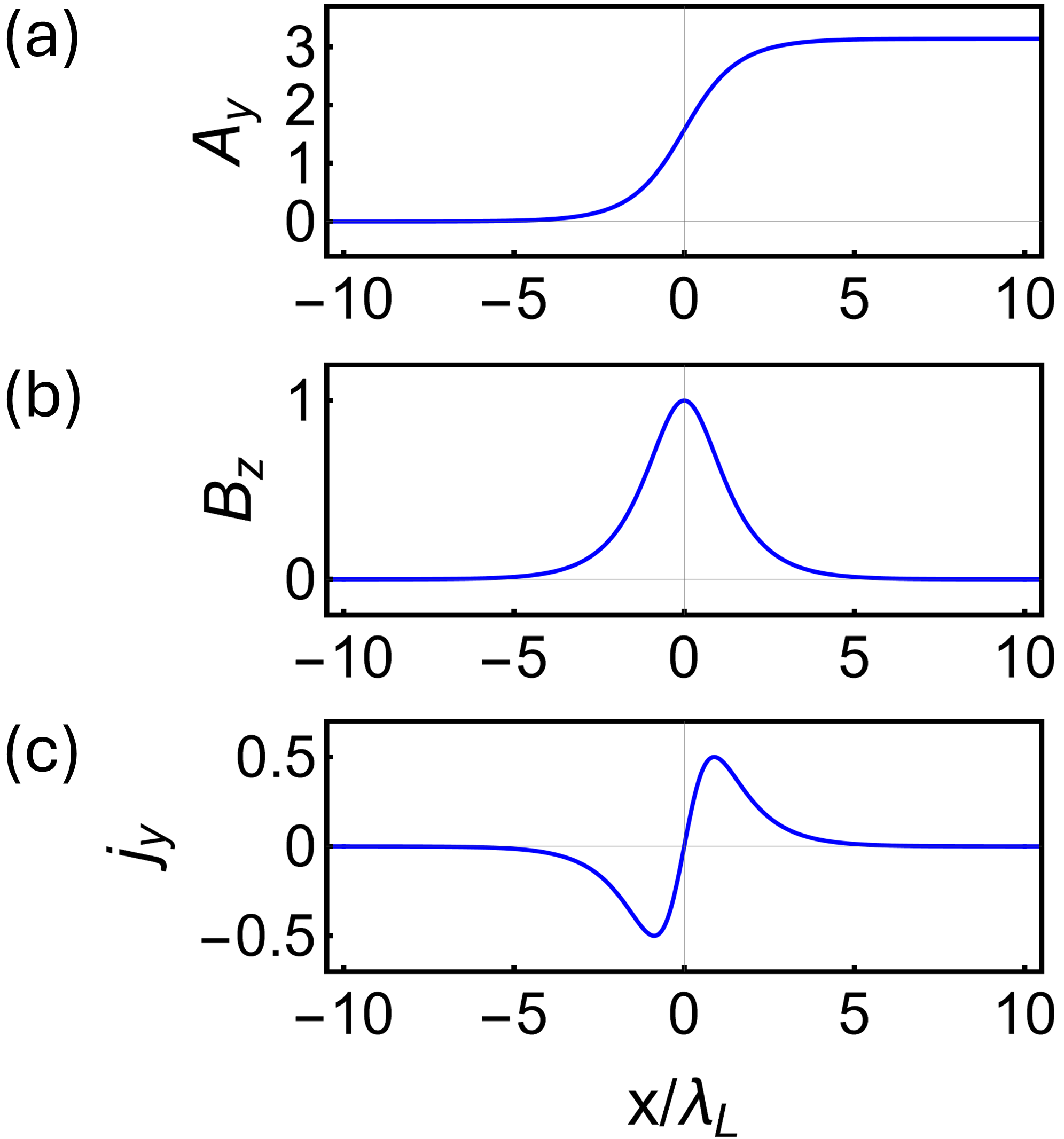}
\caption{Plots of $A_y$, $B_z$, and $j_y$ (in units of $\hbar/ea$, $\hbar/ea\lambda_L$, and $\hbar/\mu_0ea\lambda_L^2$, respectively) of the kink soliton mode along $x$ direction (Eq.~\eqref{eq:kink}, with charge $+1$). (a) The vector potential component $A_y$ has a jump of $\pi$ from $x=-\infty$ to $\infty$. (b) The region where $B_z\neq0$ forms a wall. (c) The current density $j_y$ is bidirectional on the two sides of the wall. All these profiles extend in the $yz$-plane by translation symmetry. See Fig.~\ref{fig:bulk}(c) for the 3D illustration of $B_z(x)$ and $j_y(x)$.}
\label{fig:kink}
\end{figure}

The features discussed above agree qualitatively with the free energy calculated using the mean-field approximation along high-symmetry lines for certain 2D flat-band lattice models~\cite{jiang2023pdw,sun2025flat,dunbrack2026trsbreak}. In particular, we justify that a flat-band superconductor can have a globally negative $f_s(\bd{A})$ in Appendix~\ref{app:nesting}, and illustrate it with some examples in Appendix~\ref{app:lattice}. For these models, $f_s(A_y)$ often contains higher-order harmonic terms, affecting our results only quantitatively. Combining Eqs.~\eqref{eq:maxwell2} and~\eqref{eq:cosine} leads to the time-independent sine-Gordon equation for $A_y$,
\beq\label{eq:sge}
A_y''(x)-2\mu_0\delta_2\frac{ea}{\hbar}\sin\left(\frac{2ea}{\hbar}A_y(x)\right)=0,
\eeq
which admits two different forms of real solutions, the kink and breather soliton modes~\cite{kuplevakhsky2006static}. Solitons have been connected to superconductivity in various ways, including Josephson junctions~\cite{fulton1973the,galpern1982soliton} and phase slip states~\cite{tanaka2001soliton,gurevich2003interband,kuplevakhsky2011,garaud2011topo,lin2012phase}. In these scenarios, solitons arise from the periodicity of the systems in the phase of the order parameter. In contrast, in our case, it is the \textit{phase gradient, or the vector potential}, that satisfies the sine-Gordon equation.

\begin{figure}[t!]
\centering
\includegraphics[width=0.3\textwidth]{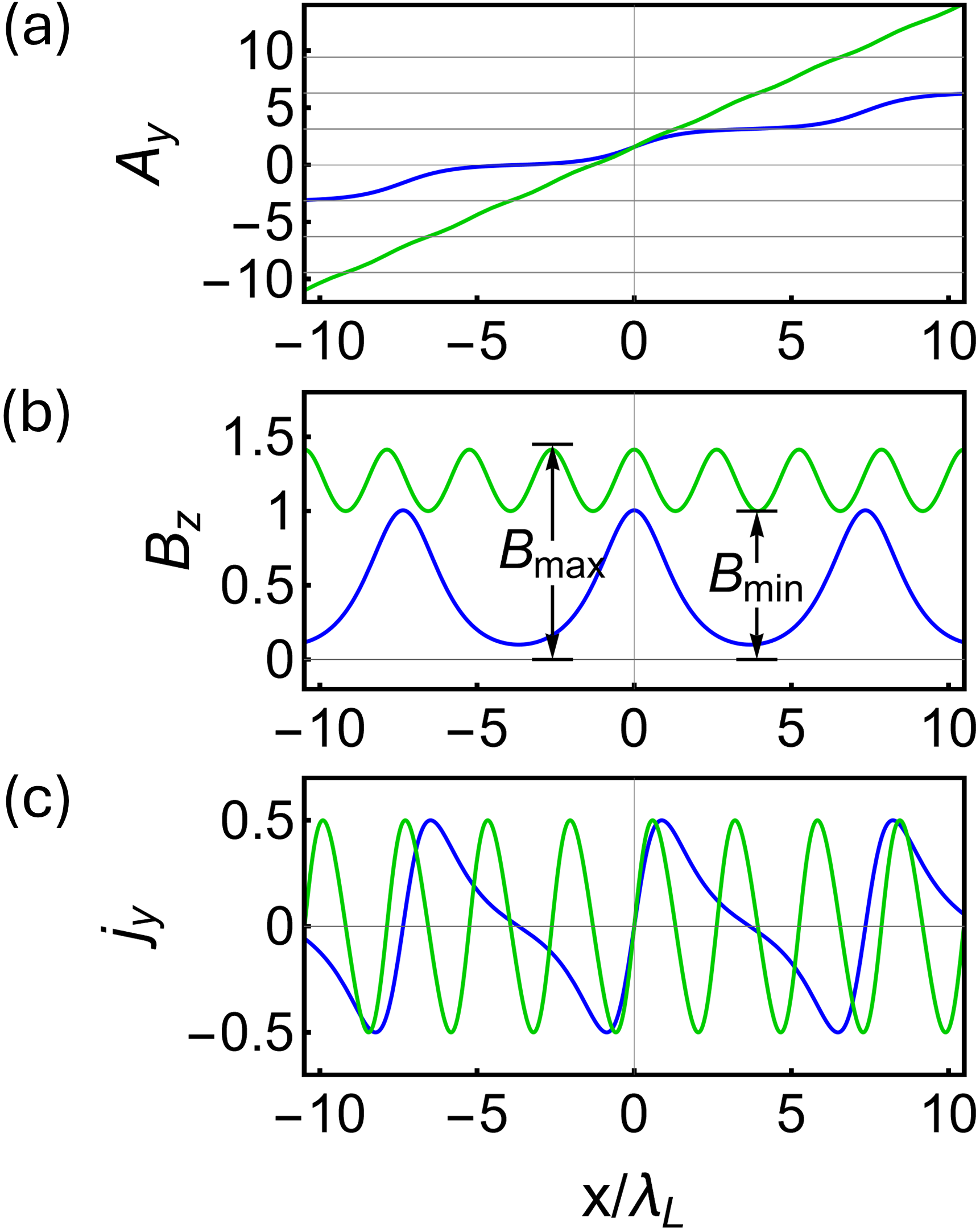}
\caption{Plots of $A_y$, $B_z$, and $j_y$ (in units of $\hbar/ea$, $\hbar/ea\lambda_L$, and $\hbar/\mu_0ea\lambda_L^2$, respectively) of the breather soliton modes along $x$ direction (Eq.~\eqref{eq:breather}, with charge $+1$) for $\alpha=0.99$ (blue) and 0.5 (green). The gray horizontal lines in (a) indicate the $\pi$ increase steps of $A_y$. The maximum and minimum $B$ fields, $B_\text{max}$ and $B_\text{min}$, are indicated in (b) for the $\alpha=0.5$ curve.}
\label{fig:breather}
\end{figure}

\section{The kink soliton solution and lower critical field}
The kink soliton solution has the form (see Appendix~\ref{app:sge})
\beq\label{eq:kink}
A_y(x)=\pm\frac{\hbar}{ea}\bigg[\sin^{-1}\tanh\bigg(\frac{x}{\lambda_L}\bigg)+\frac{\pi}{2}\bigg],
\eeq
where the $\pm$ sign is the soliton charge, indicating whether $A_y$ increases or decreases by a step of $\pi\hbar/ea$ in the positive $x$ direction, and $\lambda_L=\hbar/2ea\sqrt{\mu_0\delta_2}$ is the penetration depth along the high-symmetry direction. The charge $+1$ case is plotted in Fig.~\ref{fig:kink}(a). The derivative $A_y'(x)=B_z(x)$ vanishes at the two TRIMs that the solution connects, which we choose as $A_y=0$ and $\pi\hbar/ea$, indicating that these global minima of free energy in $\bd{A}$-space correspond to a superconducting state with \tit{zero} $B$ field (hereafter referred to as the ``zero-field state").

Then, $B_z(x)$ and $j_y(x)$ can be calculated from the derivatives of Eq.~\eqref{eq:kink}, which are proportional to $\text{sech}(x/\lambda_L)$ and $\text{sech}(x/\lambda_L)\tanh(x/\lambda_L)$, respectively, and are plotted in Figs.~\ref{fig:kink}(b) and (c). We note that a wall where $B_z\neq 0$ extends in the $yz$ plane (see Fig.~\ref{fig:bulk}(c)), and $A_y$, $B_z$, and $j_y$ exponentially decay on its both sides, with a decay length $\lambda_L$. At the center of the wall, $f_s$ reaches its maximum value, $f_{s,\text{max}}=-\delta_1+\delta_2<0$, so the material stays superconducting in the entire wall region, unlike in vortex sheets proposed by Landau and Lifshitz in superfluid helium \cite{LandauLifshitz1955,volovik2003universe} (see also Appendix~\ref{app:fluxquantization}). The bidirectional distribution of the current $j_y$ in the wall forms a pair of tightly bound antiparallel current planes.

Similar to the vortices in type-II superconductors, the walls are unstable until a magnetic field is applied. From thermodynamics~\cite{landau}, the free energy functional $F_s^H=\int \derd^3r f_s^H(\bd{r})$, with
\beq\label{eq:fsh}
f_s^H(\bd{r})=f_s(\bd{A}(\bd{r}))+\frac{1}{2\mu_0}\bd{B}(\bd{r})^2-\bd{H}\cdot\bd{B}(\bd{r}),
\eeq
attains a minimum in an applied $H$ field. Only when the applied field is above a threshold, the lower critical field $H_{c1,w}$, the last term of Eq.~\eqref{eq:fsh} overcomes the energy cost of its first two terms, stabilizing one of the two kink modes.

The lower critical field can be solved by equating $F_s^H$ of the kink solution to that of the zero-field state (see Appendix~\ref{app:hc1}). We find that for the cosine model,
\beq\label{eq:hc1}
H_{c1,w}=\frac{2}{\pi}\frac{\hbar}{\mu_0ea\lambda_L}\approx\frac{0.2\phi_0}{\mu_0a\lambda_L},
\eeq
where $\phi_0=\pi\hbar/e$ is the magnetic flux quantum. Once $H$ is above $H_{c1,w}$, multiple kinks can show up, as long as their separations are much larger than $\lambda_L$ (see Fig.~\ref{fig:bulk}(c)).

\begin{figure}[t!]
\centering
\includegraphics[width=0.48\textwidth]{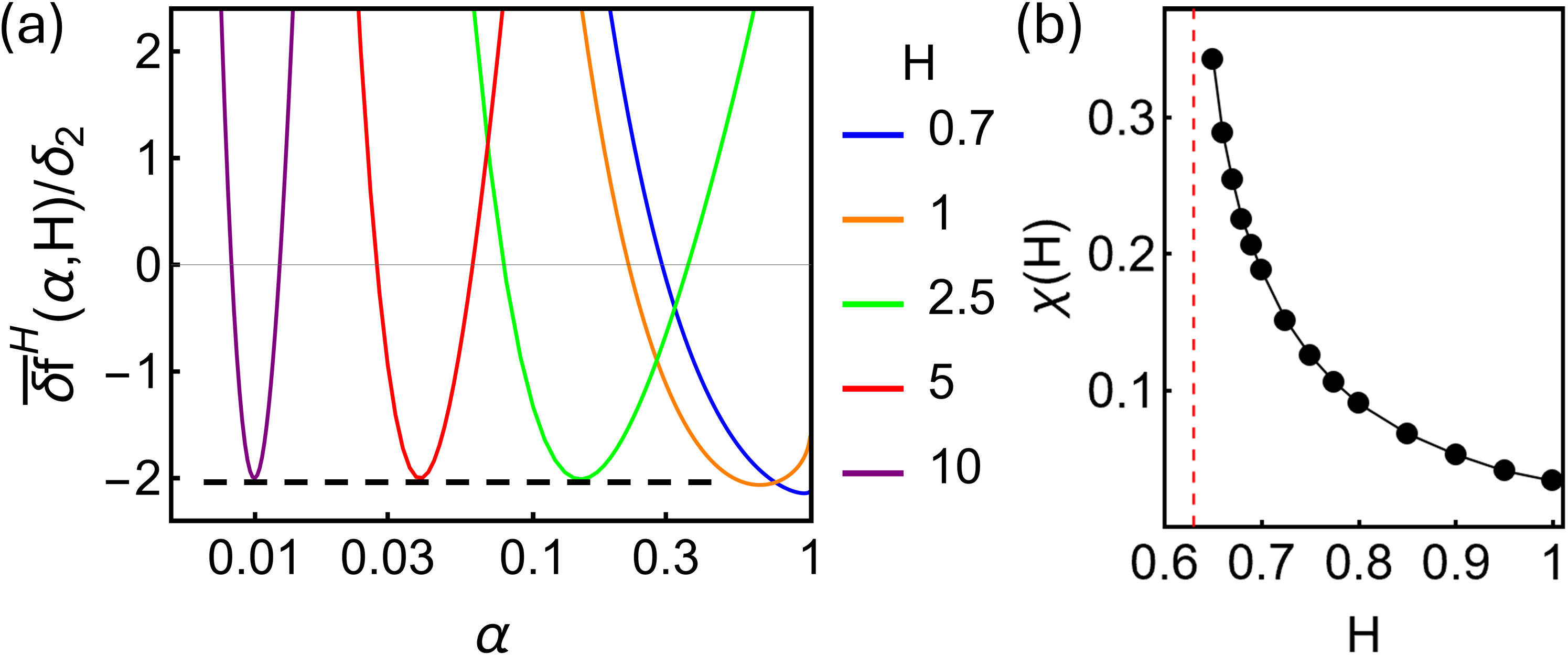}
\caption{(a) Log-linear plot of the free energy density difference between the breathers and the normal state at several $H$ fields (different colors), with $H$ in units of $\hbar/\mu_0ea\lambda_L$. The ratio $\delta_1/\delta_2$ is set to 2 for plotting. The dashed line indicates that the minima of the curves approach $-\delta_1$. (b) Magnetic susceptibility of the breathers at small fields $H\gtrsim H_{c1,w}=2/\pi\approx0.63$ (see Eq.~\eqref{eq:hc1}). The value of $H_{c1,w}$ is indicated by the red dashed line.}
\label{fig:magnetize}
\end{figure}

\begin{figure*}[t!]
\centering
\includegraphics[width=0.95\textwidth]{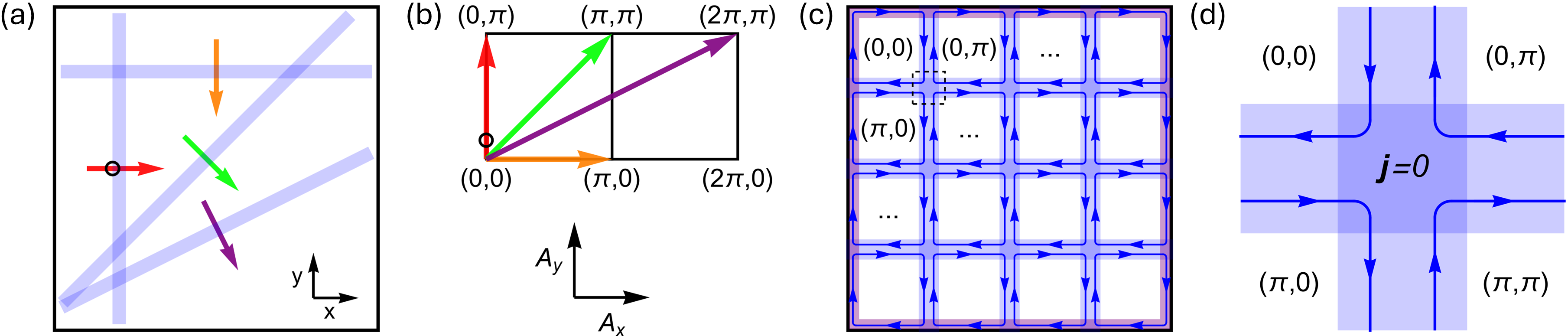}
\caption{(a) Different possible directions of walls in the real space (represented by blue ribbons) with transverse arrows. (b) The arrow associated with each wall in (a), when mapped to the $\bd{A}$-space (in the same color) by its kink solution, $\bd{A}(\bd{r})$, must connect two TRIM vector potentials (in units of $\hbar/ea$). The square lattice case is shown. The circles in (a) and (b) indicate the correspondence between an $\bd{r}$ point and an $\bd{A}$ point on the arrow. (c) A grid formed by three horizontal and vertical walls extending to the boundary of the sample (bird's-eye view of the bulk in Fig.~\ref{fig:bulk}(c)). Grid cells are labeled by the quantized values of $\bd{A}$ inside. Blue arrows indicate the directions of the current density $\bd{j}$. The pink regions near the boundary represent the usual Meissner screening regions. (d) Zoomed-in picture of the wall intersection region (dashed square) in (c). The zone center $(\pi/2,\pi/2)$ in (b) is often an extremum of $f_s(\bd{A})$, leading to $\bd{j}=0$ near the center in (d).}
\label{fig:stripes}
\end{figure*}

\section{Walls at high fields, breather solitons}
As the applied magnetic field gets stronger, the kinks tend to increase their density and even overlap to absorb more magnetic flux. To understand their evolution with the field, we invoke the breather soliton solution to Eq.~\eqref{eq:sge} (see Appendix~\ref{app:sge}):
\beq\label{eq:breather}
A_y(x)=\pm\frac{\hbar}{ea}\bigg[\text{am}\left(\alpha^{-1/2}\frac{x}{\lambda_L}\bigg|\alpha\right)+\frac{\pi}{2}\bigg],
\eeq
where $\text{am}$ is the Jacobi amplitude function, with its second argument, $\alpha$, related to the minimum $B$ field between the walls, $B_\text{min}$, through
\beq\label{eq:1overalpha}
\frac{1}{\alpha}=1+\frac{B_\text{min}^2}{4\mu_0\delta_2}.
\eeq
In Eq.~\eqref{eq:breather}, the $\pm$ sign in the solution is still the soliton charge, which is determined by $\sgn(B_\text{min})$.

With Eq.~\eqref{eq:breather}, $B_z(x)$ and $j_y(x)$ are then found to be proportional to other Jacobi elliptic functions, $\text{dn}(\alpha^{-1/2}\frac{x}{\lambda_L}|\alpha)$, and $\text{cn}(\alpha^{-1/2}\frac{x}{\lambda_L}|\alpha)\text{sn}(\alpha^{-1/2}\frac{x}{\lambda_L}|\alpha)$, respectively. We plot $A_y$, $B_z$ and $j_y$ for two $\alpha$ values, $0.99$ and $0.5$ in Fig.~\ref{fig:breather}. These profiles oscillate with the period
\beq\label{eq:period}
\Delta x=\frac{\hbar}{ea}\sqrt{\frac{\alpha}{\mu_0\delta_2}}K(\alpha)=2\alpha^{1/2}K(\alpha)\lambda_L,
\eeq
where $K(\alpha)$ is the complete elliptic integral of the first kind, which diverges at $\alpha=1$. We note that the kink solution, Eq.~\eqref{eq:kink}, can be viewed as the breather one in the $\alpha\rightarrow1$ limit, such that $\Delta x\rightarrow\infty$. As $\alpha$ decreases from 1, the period $\Delta x$ sharply decreases to finite values.

The nonzero derivative, $A_y'(x)=B_\text{min}$, of Eq.~\eqref{eq:breather} at the TRIM $A_y$ values indicates that these free energy minima correspond to a superconducting state with \tit{nonzero} $B$ field. It enables the solution $A_y$ to traverse many TRIMs, as shown in Fig.~\ref{fig:breather}(a) for the charge $+1$ case. The maximum ($B_\text{max}$) and minimum ($B_\text{min}$) $B$ fields in the mode are indicated in Fig.~\ref{fig:breather}(b). Comparing the blue with the green curves, we find that as $\alpha$ decreases, both $B_\text{max}$ and $B_\text{min}$ increase, and finally become almost equal,
\beq\label{eq:almostequal}
B_\text{max}\approx B_\text{min}\approx 2\sqrt{\frac{\mu_0\delta_2}{\alpha}},
\eeq
and their difference, $B_\text{max}-B_\text{min}$, vanishes as $\sim\sqrt{\alpha\mu_0\delta_2}$ (see Appendix~\ref{app:evolution}). This implies that in breathers of small $\alpha$ values, the field penetrates the sample almost uniformly.

\section{Magnetic susceptibility}
The thermodynamic stability of breathers at high fields can be studied by inserting the breather solution into Eq.~\eqref{eq:fsh} and comparing it with the normal state, $f_n^H(x)=-\frac{1}{2}\mu_0H^2$. The difference between the two is
\beq
\overline{\delta f^H}(\alpha,H)\equiv \frac{1}{\Delta x}\int_0^{\Delta x}\derd x[f_s^H(x)-f_n^H(x)],
\eeq
where the average over the period $\Delta x$ allows for comparing breathers of different $\alpha$ values. In Fig.~\ref{fig:magnetize}(a), we plot $\overline{\delta f^H}(\alpha,H)$ for several fields $H>H_{c1,w}$. At fixed $H$, the system seeks the $\alpha$ value to minimize $\overline{\delta f^H}$, corresponding to finding the minima of the curves. At large fields, these minima approach a constant value, namely the average free energy density of the model, $-\delta_1$ (see the dashed line in Fig.~\ref{fig:magnetize}(a)).

Reading off the $\alpha$ values of the minima of each $H$ curve in Fig.~\ref{fig:magnetize}(a), we then compute the average magnetic susceptibility of the minimal breather,
\beq
\chi(H)=1-\frac{1}{\mu_0H}\frac{1}{\Delta x}\int_0^{\Delta x}\derd xB_z(x).
\eeq
In particular, we plot $\chi(H)$ for small fields $H\gtrsim H_{c1,w}$ in Fig.~\ref{fig:magnetize}(b). It quickly drops to zero as the breather is formed. That is, the system loses its diamagnetism while preserving the superconducting gap.

The finding that $\overline{\delta f^H}$ of the minimal breather never reaches zero implies an absence of the upper critical field for the wall phase within the cosine model. Nevertheless, as the field increases and the breathers become denser, their spacing eventually becomes comparable to the lattice scale. Then, additional ingredients need to be appended. One possibility is an additional energy penalty arising from the gap amplitude gradients, i.e., terms resembling those in more microscopic models that are proportional to the coherence lengths. We note that flat-band superconductors typically have a small coherence length (can reach the lattice scale) due to their strong-coupling nature~\cite{iskin2023extracting,chen2024ginzburg,thumin2025correlation}, which leads to a small correction to the free energy density of the minimal breathers shown in Fig.~\ref{fig:magnetize}(a).

\section{Other arrangements of walls}
With the knowledge of the dynamics of Eqs.~\eqref{eq:jdef} and~\eqref{eq:maxwell} (see Appendix~\ref{app:fixedpoint}), we explain how to determine the directions of the walls in the $xy$ plane and whether they can form geometries other than depicted in Fig.~\ref{fig:bulk}(c). Given an isolated kink wall (along any direction), one can draw a transverse arrow from one side to the other (see Fig.~\ref{fig:stripes}(a)). Since the arrow connects two regions of the zero-field state, its image in the $\bd{A}$-space mapped by the kink solution $\bd{A}(\bd{r})$ must connect two TRIM vector potentials; we indicate a few of them that fit in a square lattice in Fig.~\ref{fig:stripes}(b). Upon the action of $\nabla\times\bd{A}(\bd{r})$ in Eq.~\eqref{eq:maxwell}, the arrow is rotated by $90^\circ$ during the mapping. The above process provides an approach to determine all possible wall directions from the information of $f_s(\bd{A})$ in $\bd{A}$-space. By symmetry considerations, we envision that the walls most likely arise from the high-symmetry directions, e.g., the red, orange, and green ones in Fig.~\ref{fig:stripes}(b).

Walls along different directions can also appear together to form a grid, with $\bd{A}$ quantized at TRIM values in each cell of the grid (see Fig.~\ref{fig:stripes}(c) for an example formed by horizontal and vertical walls). Intersection between walls in general causes no singularities to the fields $\bd{A}$, $\bd{B}$, and $\bd{j}$. In the zoomed-in picture of an intersection point in Fig.~\ref{fig:stripes}(d), we show one possible current distribution that is compatible with the bidirectional current in each wall. Detailed profiles of the fields near an intersection point can be solved from Eq.~\eqref{eq:maxwell} using similar methods as those for solving two-dimensional sine-Gordon equations~\cite{argyris1991finite,bratsos2007solution}, but requires the knowledge of $f_s(\bd{A})$ in the entire Cooper-pair Brillouin zone.

\section{Competition against vortices}
Additional microscopic ingredients are also pertinent to the competition between the wall and the vortex phases. The walls have one extra dimension compared to vortices; therefore, a sparse array of walls (see Fig.~\ref{fig:bulk}(c)) can absorb much more magnetic flux, thus has a much lower energy than a vortex lattice with equal density (see Appendix~\ref{app:wallarray}). This implies that their energies are comparable only when vortices have formed a much denser lattice than the wall array. Therefore, we deduce that the walls almost always win the competition against vortices if their lower critical fields are close to each other, $H_{c1,v}\approx H_{c1,w}$.

\begin{figure}[t!]
\centering
\includegraphics[width=0.45\textwidth]{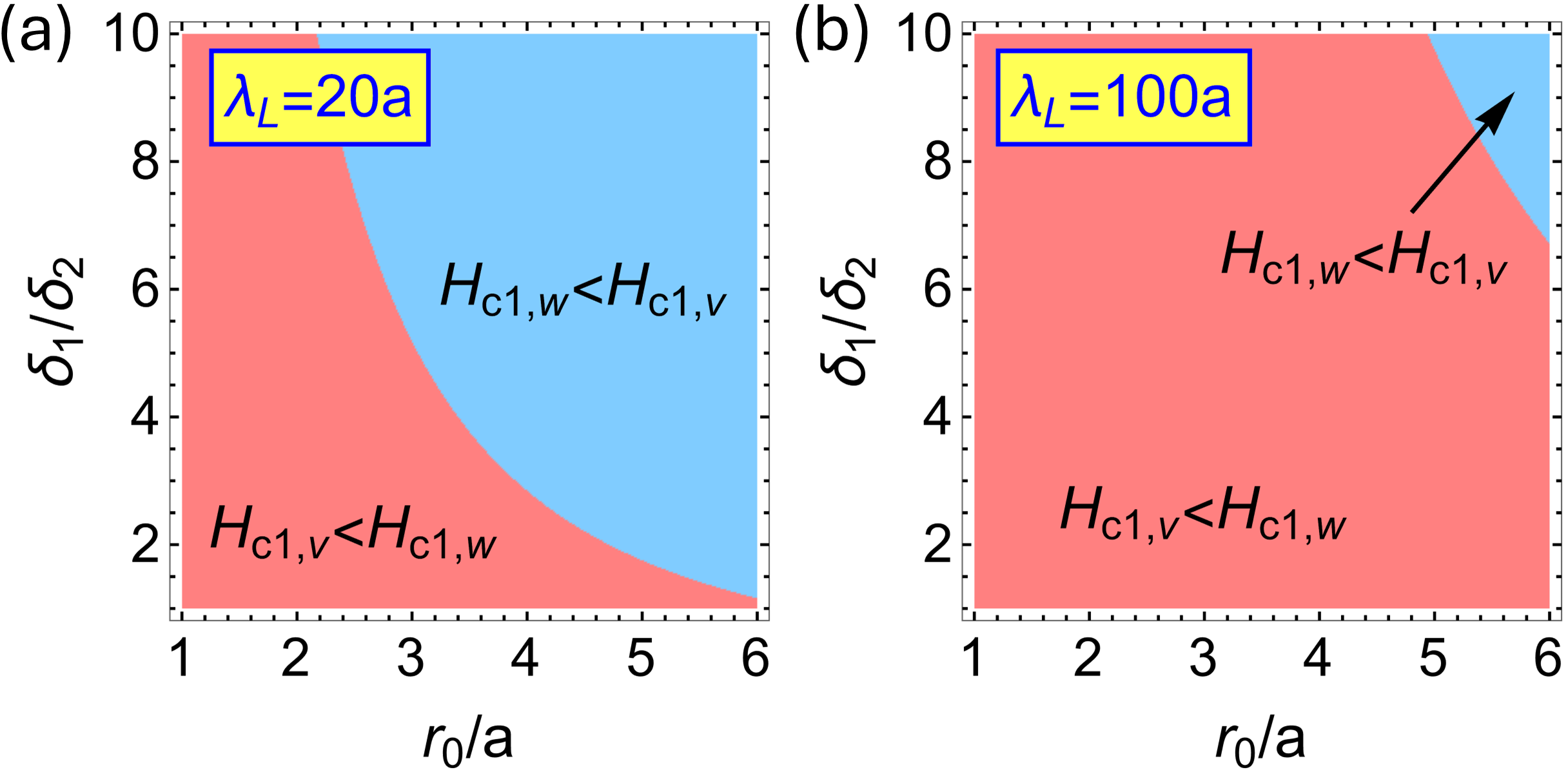}
\caption{Comparison of the lower critical fields between the wall and the vortex phases, with the blue (pink) region indicating $H_{c1,w}<H_{c1,v}$ ($H_{c1,w}>H_{c1,v}$) for small values of $\delta_1/\delta_2$ and core size $r_0\gtrsim a$. (a) is for $\lambda_L=20a$ and (b) is for $\lambda_L=100a$.}
\label{fig:diagram}
\end{figure}

Next, we compare the lower critical fields of the two phases. One can generalize the cosine model, Eq.~\eqref{eq:cosine}, to rotational symmetry to study the vortex state in flat bands (see Appendix~\ref{app:vortex} for details). We find that unlike in dispersive bands, the normal core of the vortex in flat bands constitutes a large portion of the free energy. A large normal core energy, achieved by a large core size $r_0$ and a large ratio $\delta_1/\delta_2$, can make $H_{c1,w}<H_{c1,v}$, as represented by the blue region in Fig.~\ref{fig:diagram}(a). Since the two phases have different scalings of $H_{c1}$ with $\lambda_L$: $H_{c1,w}\propto\lambda_L^{-1}$ (Eq.~\eqref{eq:hc1}) while $H_{c1,v}\propto\lambda_L^{-2}$ (see Appendix~\ref{app:vortex} for details), increasing $\lambda_L$ increases the range of parameters where $H_{c1,v}<H_{c1,w}$, as indicated by the expansion of the pink region in Fig.~\ref{fig:diagram}(b).


\section{Discussions and outlook}
We discovered a magnetic field wall phase in flat-band superconductors originating from the negativity of its free energy at any vector potential. Above its lower critical field, it allows an almost complete penetration of the field, providing a mechanism for stabilizing superconductivity in large fields. As there is a large freedom in the wall arrangements, desired configurations could be realized by purposeful design of planar defects for pinning~\cite{dolan1989vortex,prozorov2009intrinsic,kalisky2010stripes}.

Although flat-band superconductivity poses stringent constraints on the hosting materials and requires the order parameter to be comparable with the bandwidth, we showed that a negative channel of free energy along a high-symmetry direction is sufficient for the wall formation. This allows the wall phase to be found in more general band structures. Moreover, the global negativity of the free energy with the condensate momentum can be realized in other correlated states of flat bands~\cite{han2024qgn,zhang2026identifying}. Since the condition for the order parameter to be greater than the bandwidth can be achieved more easily in these states, it is a topic of future research to find whether the wall phase can be extended there. Our results suggest that flat-band superconductors can sustain high magnetic fields and form wall textures; this broadens our fundamental understanding of superconductivity and the scope of its applications.

\section*{Acknowledgments}
We thank Pertti Hakonen and Pauli Virtanen for useful discussions. This work is supported by the Research Council of Finland under project number 349313 and through the Finnish Quantum Flagship, project number 359240, by Jane and Aatos Erkko Foundation, Keele Foundation, Magnus Ehrnrooth Foundation, and a collaboration between the Kavli Foundation, Klaus Tschira Stiftung, and Kevin Wells, as part of the SuperC collaboration, and by a grant from the Simons Foundation (SFI-MPS-NFS-00006741-12, P.T.) in the Simons Collaboration on New Frontiers in Superconductivity. This work is part of the Finnish Centre of Excellence in Quantum Materials (QMAT).

\appendix

\section{Local self-consistency approximation}
\label{app:fsdensity}
\noindent(Throughout the appendices, we take $e=\hbar=a=1$ unless otherwise specified.)\\

The local self-consistency approximation ascribes the spatial dependence of the free energy density $f_s$ to the gauge-invariant vector potential $\bd{A}$ in an inhomogeneous superconductor, i.e., we write $f_s(\bd{r})\approx f_s(\bd{A}(\bd{r}))$. This approximation becomes exact when the nonlocal effect of Cooper pairs~\cite{pippard1953the} is extremely weak, i.e., when the coherence length $\xi$ and other microscopic length scales are much smaller than the penetration depth $\lambda_L$.

Within the local self-consistency approximation, the free energy as a function of $\bd{A}$ is calculated from the grand potential $\Omega$,
\beq
F_s(\bd{A})=\Omega(\bd{A},\Delta_\bd{A},\mu_\bd{A})+\mu_\bd{A}N_e,
\eeq
with $\mu_\bd{A}$ the chemical potential (which also depends on $\bd{A}$ locally) and $N_e$ the number of electrons. The grand potential is evaluated from the mean-field Hamiltonian $\hham(\bd{A},\Delta_\bd{A},\mu_\bd{A})$,
\beq
\Omega(\bd{A},\Delta_\bd{A},\mu_\bd{A})=-\frac{1}{\beta}\ln\tr\{e^{-\beta\hham(\bd{A},\Delta_\bd{A},\mu_\bd{A})}\},
\eeq
with $1/\beta=k_B T$ where $T$ is the temperature and $k_B$ the Boltzmann constant. All of the extensive quantities above, i.e., $\Omega$, $F_s$ and $N_e$, are counted in a volume $\mathcal{V}$ (much larger than the scale of $\xi$) in which $\bd{A}$ is treated as a constant. The order parameter $\Delta_\bd{A}$ and the chemical potential $\mu_\bd{A}$ are solved from the local self-consistency equations, with the vector potential $\bd{A}$ treated with Peierls substitution in the Hamiltonian $\hham$ (see Appendix~\ref{app:gauge} for further details). Then the free energy density $f_s(\bd{A})=F_s(\bd{A})/\mathcal{V}$, and the current density $\bd{j}$, as given by Eq.~\eqref{eq:jdef}, depends on $\bd{A}$ \tit{only}.

\section{Gauge invariance of free energy and supercurrent}
\label{app:gauge}
In BCS theory, the supercurrent density $\bd{j}$ is given by
\beq\label{eq:BCSjdef}
\bd{j}=-D_s\left(\bd{A}^{(\chi)}+\frac{\hbar}{2e}\nabla\varphi\right)=-D_s\bd{A},
\eeq
where $D_s$ is the superfluid weight, $\nabla\varphi$ is the gradient of the order parameter phase, while $\bd{A}^{(\chi)}$ and $\bd{A}=\bd{A}^{(\chi)}+\frac{\hbar}{2e}\nabla\varphi$ are the gauge-dependent and gauge-invariant vector potentials, respectively (the notation $\bd{A}^{(\chi)}$ comes from the gauge transformation in electrodynamics, $\bd{A}^{(\chi)}=\bd{A}+\nabla\chi$, with $\chi$ a scalar function). Therefore, $\bd{j}$ is gauge-invariant. Below, we derive the gauge invariance within the mean-field framework and show that it is still valid when $\bd{j}$ has a nonlinear relation with $\bd{A}$.
\subsection{Current operator and gauge invariance}
\label{app:currentoperator}
The vector potential $\bd{A}^{(\chi)}$ enters a tight-binding non-interacting Hamiltonian $\hham_0$ through the Peierls substitution,
\beq\label{eq:h0a}
\hham_0(\bd{A}^{(\chi)})=\sum_{i\alpha,j\beta,\sigma}t^\sigma_{i\alpha,j\beta}e^{-i\frac{e}{\hbar}\bd{A}^{(\chi)}\cdot(\bd{r}_{i\alpha}-\bd{r}_{j\beta})}c_{i\alpha\sigma}^\dagger c_{j\beta\sigma}.
\eeq
where $i,j$ label the unit cell, $\alpha,\beta$ label the orbitals inside a unit cell, and $\sigma$ labels the spin. Here $\bd{r}_{i\alpha}=\bd{R}_i+\bd{x}_\alpha$ is the position of orbital $i\alpha$, with $\bd{R}_i$, $\bd{x}_\alpha$ the position of unit cell $i$, and the position of orbital $\alpha$ with respect to the unit cell, respectively. Spin-orbit coupling is neglected, and $t^\sigma_{i\alpha,j\beta}$ is the parameter of hopping from orbital $j\beta$ to $i\alpha$ for spin $\sigma$, while $c^\dagger,c$ are creation and annihilation operators. Eq.~\eqref{eq:h0a} describes the Hamiltonian in a local region where $\bd{A}^{(\chi)}$ can be viewed as constant.

On the other hand, the phase gradient enters the interaction through mean-field decoupling. We start from an on-site density-density interaction (both intra- and interorbital interactions $U_{\alpha\beta}$ are involved)
\begin{align}
\hham_I=&-\sum_{i,\alpha\beta}U_{\alpha\beta}\hat{n}_{i\alpha\uparrow}\hat{n}_{i\beta\downarrow}\\
=&-\sum_{\bd{k}\bd{k}'\bd{q},\alpha\beta}\frac{U_{\alpha\beta}}{N_c}c_{\bd{k}+\bd{q},\alpha\uparrow}^\dagger c_{-\bd{k}+\bd{q},\beta\downarrow}^\dagger c_{-\bd{k}'+\bd{q},\beta\downarrow}c_{\bd{k}'+\bd{q},\alpha\uparrow},\nonumber
\end{align}
where
\beq
c_{\bd{k}\alpha\sigma}=\frac{1}{\sqrt{N_c}}\sum_ie^{-i\bd{k}\cdot\bd{r}_{i\alpha}}c_{i\alpha\sigma},
\eeq
with $N_c$ the number of unit cells. We assume singlet pairing, and do not require all $U_{\alpha\beta}$ to be attractive, but the results can be generalized to other pairings. Assuming that the condensate is formed by Cooper pairs $(\bd{k}+\bd{q}\uparrow,-\bd{k}+\bd{q}\downarrow)$ with fixed $\bd{q}$ ($2\bd{q}=\bd{Q}=\nabla\varphi$ is the phase gradient), mean-field decoupling leads to the reduced interaction
\beq
\hham^{\text{MF}}_I(\bd{q})=\sum_{\bd{k},\alpha\beta}(\Delta_{\alpha\beta}c_{\bd{k}+\bd{q},\alpha\uparrow}^\dagger c_{-\bd{k}+\bd{q},\beta\downarrow}^\dagger+h.c.)+N_c\sum_{\alpha\beta}\frac{|\Delta_{\alpha\beta}|^2}{U_{\alpha\beta}},
\eeq
with order parameters
\beq\label{eq:order}
\Delta_{\alpha\beta}=-\frac{U_{\alpha\beta}}{N_c}\sum_\bd{k}\la c_{-\bd{k}+\bd{q},\beta\downarrow}c_{\bd{k}+\bd{q},\alpha\uparrow}\ra_{\bd{A}^{(\chi)},\bd{q}},
\eeq
where $\la\ra_{\bd{A}^{(\chi)},\bd{q}}$ represents the average in the condensate.

Note that both $\bd{A}^{(\chi)}$ and $\bd{q}$ parameterize the total Hamiltonian
\beq\label{eq:htotal}
\hham(\bd{A}^{(\chi)},\bd{q})=\hham_0(\bd{A}^{(\chi)})+\hham^{\text{MF}}_I(\bd{q})-\mu\sum_{i\alpha\sigma}\hat{n}_{i\alpha\sigma},
\eeq
with $\mu$ the chemical potential. Then, $\Delta_{\alpha\beta}$ and $\mu$ depend on $\bd{A}^{(\chi)}$ and $\bd{q}$ implicitly through self-consistency equations. We define the total current operator $\hat{\bd{J}}$ as~\cite{mahan} ($\bd{J}=\bd{j}\mathcal{V}$, with $\mathcal{V}$ the volume)
\beq\label{eq:current}
\hat{\bd{J}}(\bd{A}^{(\chi)},\bd{q})=\frac{i}{\hbar}[\hham(\bd{A}^{(\chi)},\bd{q}),\hat{\bd{P}}],
\eeq
with $\hat{\bd{P}}=-e\sum_{i\alpha\sigma}\bd{r}_{i\alpha}c_{i\alpha\sigma}^\dagger c_{i\alpha\sigma}$ the charge polarization. This definition ensures charge conservation by itself. The two terms in $\hat{\bd{J}}$ contributed by $\hham_0(\bd{A}^{(\chi)})$ and $\hham^{\text{MF}}_I(\bd{q})$ are
\begin{align}
\hat{\bd{J}}_1=&-\frac{ie}{\hbar}\sum_{i\alpha,j\beta,\sigma}t^\sigma_{i\alpha,j\beta}e^{-i\frac{e}{\hbar}\bd{A}^{(\chi)}\cdot(\bd{r}_{i\alpha}-\bd{r}_{j\beta})}(\bd{r}_{j\beta}-\bd{r}_{i\alpha})c_{i\alpha\sigma}^\dagger c_{j\beta\sigma} \nonumber\\
=&-\frac{\derd\hham_0(\bd{A}^{(\chi)})}{\derd\bd{A}^{(\chi)}},
\end{align}
and
\begin{align}
\hat{\bd{J}}_2=&\bigg[\frac{ie}{\hbar N_c}\sum_{i\alpha\beta,\bd{k}\bd{q}'}\Delta_{\alpha\beta}(\bd{r}_{i\beta}e^{i\bd{q}'\cdot\bd{r}_{i\beta}}c_{\bd{k}+\bd{q},\alpha\uparrow}^\dagger c_{-\bd{k}+\bd{q}-\bd{q}',\beta\downarrow}^\dagger \nonumber\\
&+\bd{r}_{i\alpha}e^{i\bd{q}'\cdot\bd{r}_{i\alpha}}c_{\bd{k}+\bd{q}-\bd{q}',\alpha\uparrow}^\dagger c_{-\bd{k}+\bd{q},\beta\downarrow}^\dagger)\bigg]+h.c., \label{eq:mfcommute}
\end{align}
respectively, whereas the chemical potential term does not contribute. Using Eq.~\eqref{eq:order}, it is straightforward to verify that $\hat{\bd{J}}_2$ has zero expectation value in the condensate, therefore,
\beq\label{eq:javerage}
\la \hat{\bd{J}}\ra_{\bd{A}^{(\chi)},\bd{q}}=\la \hat{\bd{J}}_1\ra_{\bd{A}^{(\chi)},\bd{q}}=-\bigg\la\frac{\derd\hham_0(\bd{A}^{(\chi)})}{\derd\bd{A}^{(\chi)}}\bigg\ra_{\bd{A}^{(\chi)},\bd{q}}.
\eeq

With these established, we perform a gauge transformation, i.e., change of variables $\bd{A}^{(\chi)}\rightarrow \bd{A}^{(\chi)}-(\hbar/e)\delta\bd{q}$, $\bd{q}\rightarrow\bd{q}+\delta\bd{q}$ to $\hham(\bd{A}^{(\chi)},\bd{q})$. This transformation amounts to replacing the basis $c_{\bd{k}\alpha\sigma}\rightarrow\tilde{c}_{\bd{k}\alpha\sigma}\equiv c_{\bd{k}+\delta\bd{q},\alpha\sigma}$ while keeping the variables unchanged in these second-quantized operators, i.e.,
\begin{align}
\label{eq:basischange}
&\hham(\bd{A}^{(\chi)}-\frac{\hbar}{e}\delta\bd{q},\bd{q}+\delta\bd{q})=\tilde{\hham}(\bd{A}^{(\chi)},\bd{q}), \nonumber\\
&\hat{\bd{J}}(\bd{A}^{(\chi)}-\frac{\hbar}{e}\delta\bd{q},\bd{q}+\delta\bd{q})=\tilde{\hat{\bd{J}}}(\bd{A}^{(\chi)},\bd{q}),
\end{align}
where $\tilde{\hham},\tilde{\hat{\bd{J}}}$ refer to the same operators as $\hham,\hat{\bd{J}}$ but with new basis $\tilde{c}$. Then, the expectation values of these operators are invariants, e.g., $\bd{J}(\bd{A}^{(\chi)},\bd{q})\equiv\la \hat{\bd{J}}(\bd{A}^{(\chi)},\bd{q})\ra_{\bd{A}^{(\chi)},\bd{q}}$ satisfies
\beq
\bd{J}(\bd{A}^{(\chi)}-\frac{\hbar}{e}\delta\bd{q},\bd{q}+\delta\bd{q})=\bd{J}(\bd{A}^{(\chi)},\bd{q}).
\eeq
Therefore, we conclude that the expectation values must be a function of $\bd{A}=\bd{A}^{(\chi)}+\frac{\hbar}{e}\bd{q}$. It is noteworthy that Eq.~\eqref{eq:order}, and the other self-consistency equation determining $\mu$ (the number equation),
\beq\label{eq:mueq}
\sum_{i\alpha\sigma}\la\hat{n}_{i\alpha\sigma}\ra_{\bd{A}^{(\chi)},\bd{q}}=N_e ,
\eeq
with $N_e$ the average number of electrons in the condensate, are both in terms of expectation values, therefore, $\Delta_{\alpha\beta}$ and $\mu$ are also functions of $\bd{A}$, and we denote them as $\Delta_{\bd{A},\alpha\beta}$ and $\mu_{\bd{A}}$.

We emphasize that the gauge-invariance derived here does not rely on a specific relation between $\bd{J}$ and $\bd{A}$. We do not seek an expansion of $\bd{A}^{(\chi)}$ in the Peierls substitution coefficient $e^{-i\frac{e}{\hbar}\bd{A}^{(\chi)}\cdot(\bd{r}_{i\alpha}-\bd{r}_{j\beta})}$, so we lose no higher order terms. Therefore, the gauge invariance applies to our scenarios where $\bd{A}$ can reach the order of magnitude of the reciprocal lattice, $1/a$.

We also note that the superconductor cannot stay in a state with a constant vector potential $\bd{A}_0$ and nonzero current density $\bd{j}=\bd{j}_{\bd{A}_0}\neq0$ in the entire space. The reason is that such a state simultaneously requires the $B$ field to be spatial dependent (since $\nabla\times\bd{B}=\mu_0\bd{j}$), and $\bd{B}=0$ (since $\nabla\times\bd{A}_0=0$), a contradiction. Therefore, in the region where $\bd{j}\neq 0$, both $\bd{A}$ and $\bd{j}$ must vary in space. Throughout the derivation of the gauge-invariance above, we implicitly use the local self-consistency assumption that $\bd{A}$ can be locally treated as a constant, and the local self-consistency is established. This assumption is equivalent to stating that $\lambda_L\gg\xi$. Therefore, the results that the local physical quantities $\bd{j}$, $\Delta$, and $\mu$ are functions of a single variable $\bd{A}$ are only valid in this limit. As $\xi$ becomes large, $\bd{A}$ is no longer directly associated with these quantities.
\subsection{Relation between current and free energy}
To calculate the current from the free energy, it is sufficient to consider a condensate with $\bd{q}=0$ and vector potential $\bd{A}$. Since $\Delta$ and $\mu$ depend on $\bd{A}$ implicitly, we write Eq.~\eqref{eq:htotal} as $\hham(\bd{A},\Delta_\bd{A},\mu_\bd{A})$. Then, Eq.~\eqref{eq:javerage} becomes
\begin{align}
\bd{J}(\bd{A})=&-\bigg\la\frac{\partial\hham}{\partial\bd{A}}\bigg|_{\Delta_\bd{A},\mu_\bd{A}}\bigg\ra \nonumber\\
=&-\frac{1}{\Xi}\tr\bigg\{\frac{\partial\hham}{\partial\bd{A}}\bigg|_{\Delta_\bd{A},\mu_\bd{A}}e^{-\beta\hham}\bigg\}=-\frac{\partial\Omega}{\partial\bd{A}}\bigg|_{\Delta_\bd{A},\mu_\bd{A}},
\end{align}
where $\Xi=\tr\{e^{-\beta\hham}\}$ is the grand partition function and
\beq\label{eq:grandpotential}
\Omega(\bd{A},\Delta_\bd{A},\mu_\bd{A})=-\frac{1}{\beta}\ln\tr\{e^{-\beta\hham(\bd{A},\Delta_\bd{A},\mu_\bd{A})}\}
\eeq
is the grand potential. In Eq.~\eqref{eq:javerage}, the current operator is expressed as the total derivative of the noninteracting $\hham_0$ with respect to $\bd{A}$. Since the interacting part $\hham_I^\text{MF}$ depends on $\bd{A}$ implicitly via $\Delta_{\bd{A}}$ and $\mu_\bd{A}$, here it becomes the partial derivative of $\hham$ with $\bd{A}$.

The Helmholtz free energy of the grand canonical ensemble is $F_s(\bd{A})\equiv\Omega(\bd{A},\Delta_\bd{A},\mu_\bd{A})+\mu_\bd{A}N_e$, and we find
\begin{align}
\frac{\derd F_s}{\derd\bd{A}}=&\frac{\partial \Omega}{\partial\bd{A}}\bigg|_{\Delta_\bd{A},\mu_\bd{A}}+\frac{\partial \Omega}{\partial\Delta_\bd{A}}\bigg|_{\bd{A},\mu_\bd{A}}\frac{\derd\Delta_\bd{A}}{\derd\bd{A}} \nonumber\\
&+\bigg(\frac{\partial \Omega}{\partial\mu_\bd{A}}\bigg|_{\bd{A},\Delta_\bd{A}}+N_e\bigg)\frac{\derd\mu_\bd{A}}{\derd\bd{A}}=\frac{\partial \Omega}{\partial\bd{A}}\bigg|_{\Delta_\bd{A},\mu_\bd{A}},
\end{align}
where the second and third terms cancel by Eq.~\eqref{eq:order} and~\eqref{eq:mueq}, respectively. Then we can express the current as the total derivative of free energy,
\beq
\bd{J}(\bd{A})=-\frac{\derd F_s}{\derd\bd{A}}.
\eeq
\section{Free energy of flat-band superconductors at nonzero vector potential}
\label{app:nesting}
Two properties of flat-band superconductors that distinguish them from conventional ones are discussed below.\\

\tit{(i) At low temperatures and near half filling, a flat-band superconductor has negative free energy at a vector potential $\bd{A}$ if the order parameter at $\bd{A}$ is nonzero.}\\

The opposite direction of this statement is trivial, since whenever the order parameters vanish, the superconducting state becomes normal and $f_s(\bd{A})=0$. To show (i), we consider a single flat band $m$ that is isolated from other bands by a gap larger than the superconducting gap. However, the results below can be generalized to the cases of multiple flat bands if they are nearly degenerate.

We assume a condensate with $\bd{q}=0$ and a fixed $\bd{A}$ (since the free energy is gauge-invariant). We project the Hamiltonian $\hham(\bd{A},0)$ of Eq.~\eqref{eq:htotal} to the flat band to get~\cite{jiang2023pdw}
\beq\label{eq:hband}
\begin{split}
\hham^{(m)}(\bd{A},0)=&\sum_\bd{k}\bd{d}^\dagger_{m\bd{k}}(\bd{A})\ham^{(m)}_{\text{BdG},\bd{k}}(\bd{A})\bd{d}_{m\bd{k}}(\bd{A})\\
&+\sum_\bd{k}\xi_{m,\bd{k}-\bd{A}}+N_c\sum_{\alpha\beta}\frac{|\Delta_{\bd{A},\alpha\beta}|^2}{U_{\alpha\beta}}.
\end{split}
\eeq
Here, time-reversal symmetry is imposed, implying that the band dispersion $\varepsilon^\uparrow_{m,\bd{k}}=\varepsilon^\downarrow_{m,-\bd{k}}$. We define $\xi_{m,\bd{k}}\equiv \varepsilon^\uparrow_{m,\bd{k}}-\mu_\bd{A}$.
For perfectly flat bands, $\varepsilon^\sigma_{m,\bd{k}}\equiv0$, while $\mu_\bd{A}$ can be of the order of $\Delta_{\bd{A},\alpha\beta}$, depending on the band filling. The Nambu spinor in the band basis,
\beq
\bd{d}_{m\bd{k}}(\bd{A})=\begin{pmatrix}
&\sum_\alpha\psi^*_{m,\bd{k}+\bd{A},\alpha}c_{\bd{k}\alpha\uparrow}\\
&\sum_\alpha\psi^*_{m,\bd{k}-\bd{A},\alpha}c^\dagger_{-\bd{k}\alpha\downarrow}
\end{pmatrix},
\eeq
with $\psi_{m,\bd{k},\alpha}$ the Bloch state eigenvector of the spin-$\uparrow$ band, and
\beq
\ham^{(m)}_{\text{BdG},\bd{k}}(\bd{A})=\begin{pmatrix}
\xi_{m,\bd{k}+\bd{A}} & \Delta_{m,\bd{k}}(\bd{A})\\
\Delta_{m,\bd{k}}(\bd{A})^* & -\xi_{m,\bd{k}-\bd{A}}
\end{pmatrix}
\eeq
is the band-projected BdG matrix, with
\beq\label{eq:gapband}
\Delta_{m,\bd{k}}(\bd{A})\equiv\sum_{\alpha\beta}\psi_{m,\bd{k}+\bd{A},\alpha}^*\Delta_{\bd{A},\alpha\beta}\psi_{m,\bd{k}-\bd{A},\beta}.
\eeq
Diagonalizing Eq.~\eqref{eq:hband} and computing the free energy of the grand canonical ensemble, we obtain
\begin{align}
F_s(\bd{A})=&-\frac{1}{\beta}\sum_\bd{k}\ln[(1+e^{-\beta E_{m,\bd{k},+}(\bd{A})})(1+e^{-\beta E_{m,\bd{k},-}(\bd{A})})] \nonumber\\
&+\sum_\bd{k}\xi_{m,\bd{k}-\bd{A}}+N_c\sum_{\alpha\beta}\frac{|\Delta_{\bd{A},\alpha\beta}|^2}{U_{\alpha\beta}}+\mu_\bd{A}N_e,
\end{align}
where $E_{m,\bd{k},\pm}(\bd{A})$ are the two eigenvalues of $\ham^{(m)}_{\text{BdG},\bd{k}}(\bd{A})$.

We now focus on the ideally flat-band, half-filling, zero-temperature limit. In this limit, $\mu_\bd{A}\rightarrow0$ for all $\bd{A}$, therefore, $\xi_{m\bd{k}}\rightarrow 0$ and $E_{m,\bd{k},\pm}(\bd{A})\rightarrow \pm|\Delta_{m,\bd{k}}(\bd{A})|$. We find
\beq\label{eq:fa0}
F(\bd{A})=-\sum_\bd{k}|\Delta_{m,\bd{k}}(\bd{A})|+N_c\sum_{\alpha\beta}\frac{|\Delta_{\bd{A},\alpha\beta}|^2}{U_{\alpha\beta}}.
\eeq
This expression can be further simplified using the gap equation. In this limit, the gap Eq.~\eqref{eq:order} reads
\beq\label{eq:gap1}
\Delta_{\bd{A},\alpha\beta}=\frac{U_{\alpha\beta}}{2N_c}\sum_\bd{k}\psi_{m,\bd{k}-\bd{A},\beta}^*\psi_{m,\bd{k}+\bd{A},\alpha}e^{i\text{Arg}\{\Delta_{m,\bd{k}}(\bd{A})\}}.
\eeq
Manipulating this with Eq.~\eqref{eq:fa0}, then, the second term of Eq.~\eqref{eq:fa0} exactly cancels half of its first term, leading to
\beq\label{eq:fa}
F_s(\bd{A})=-\frac{1}{2}\sum_\bd{k}|\Delta_{m,\bd{k}}(\bd{A})|.
\eeq

Equation~\eqref{eq:fa} states that $|\Delta_{m,\bd{k}}(\bd{A})|$ is exactly twice of the condensation energy per $\bd{k}$. For a given $\bd{A}$, whenever the gap equation has nonzero solutions, $\Delta_{\bd{A},\alpha\beta}$ is a nonzero matrix. Caution is needed for that $|\Delta_{m,\bd{k}}(\bd{A})|$ may vanish at some $\bd{k}$-points even if $\Delta_{\bd{A},\alpha\beta}$ is a nonzero matrix, but in general, it cannot vanish at every $\bd{k}$-point. Thus, we conclude $F_s(\bd{A})<0$ whenever the order parameters are nonzero.

\tit{Comparison with BCS superconductors in dispersive bands.} The condensation energy of a superconductor in dispersive bands can deplete at a vector potential $\bd{A}$ even if $\Delta_\bd{A}\neq0$, i.e., through first-order transitions. This happens typically at temperatures well below $T_c$. In this case, the free energy density is approximately $f_s(\bd{A})\approx f_s(0)+\frac{1}{2}D_sA^2$ if the depairing effect is neglected, where the phase gradient term $\frac{1}{2}D_sA^2$ drives $f_s$ to zero at $A_c\approx\sqrt{2|f_s(0)|/D_s}$, whereas the order parameter $\Delta_\bd{A}$ may vanish at a much larger vector potential (the depairing momentum). Therefore, (i) simply tells us that for flat bands, the phase gradient cannot deplete the condensation energy in the low-temperature, half-filling limit; the superconducting-normal transition driven by depairing, if it exists, must be second-order.\\

\tit{(ii) For flat bands at low temperatures and near half filling, the solutions to the gap equation at different $\bd{A}$ only differ quantitatively.}\\

This statement means that if the pairing mechanism exists in the orbital level, then the gap equation, in general, has nonzero solutions at all $\bd{A}$ and the order parameters at different $\bd{A}$ only differ by magnitude. We provide the arguments below that lead to this observation or assumption.

Instead of analyzing the magnitude of each entry of $\Delta_{\bd{A},\alpha\beta}$, we multiply Eq.~\eqref{eq:gap1} with $\psi_{m,\bd{k}+\bd{A},\alpha}^*\psi_{m,\bd{k}-\bd{A},\beta}$ and sum over $\alpha,\beta$ to get a self-consistency equation for $\Delta_{m,\bd{k}}(\bd{A})$:
\beq\label{eq:gap2}
\Delta_{m,\bd{k}}(\bd{A})=\sum_{\bd{k}'}O(\bd{k},\bd{k}',\bd{A})e^{i\text{Arg}\{\Delta_{m,\bd{k}'}(\bd{A})\}},
\eeq
where
\beq
O(\bd{k},\bd{k}',\bd{A})=\sum_{\alpha\beta}\frac{U_{\alpha\beta}}{2N_c}\psi_{m,\bd{k}'-\bd{A},\beta}^*\psi_{m,\bd{k}'+\bd{A},\alpha}\psi_{m,\bd{k}+\bd{A},\alpha}^*\psi_{m,\bd{k}-\bd{A},\beta}.
\eeq
The r.h.s. of Eq.~\eqref{eq:gap2} can be viewed as a summation of vectors in the complex plane. We notice that the region of $\bd{k}'\approx \bd{k}$ always contribute constructively, since in this region the phase $e^{i\text{Arg}\{\Delta_{m\bd{k}'}(\bd{A})\}}$ is close to that of $\Delta_{m\bd{k}}(\bd{A})$, while
\beq\label{eq:onew}
O(\bd{k},\bd{k}',\bd{A})\approx\sum_{\alpha\beta}\frac{U_{\alpha\beta}}{2N_c}|\psi_{m,\bd{k}+\bd{A},\alpha}|^2|\psi_{m,\bd{k}-\bd{A},\beta}|^2
\eeq
is real. The regions of $\bd{k}'$ that are far from $\bd{k}$ instead cancel or contribute only partially. Equation~\eqref{eq:onew} also tells us that if there is a channel with large attractive interaction, $U_{\alpha\beta}>0$, dominating over other repulsive ones, then $O(\bd{k},\bd{k},\bd{A})>0$ and Eq.~\eqref{eq:gap2} is very likely to have nonzero solutions. This result can be qualitatively stated as
\beq\label{eq:uab}
|\Delta_{m,\bd{k}}(\bd{A})|\propto U_{\alpha\beta}|\psi_{m,\bd{k}+\bd{A},\alpha}|^2|\psi_{m,\bd{k}-\bd{A},\beta}|^2,
\eeq
with $U_{\alpha\beta}>0$ the leading attractive interaction.

As discussed previously, since $|\Delta_{m,\bd{k}}(\bd{A})|$ may vanish at some $\bd{k}$-points, a better measure of the magnitude of the set of order parameters, $\{\Delta_{\bd{A},\alpha\beta}\}$, is to average Eq.~\eqref{eq:uab} over the entire Brillouin zone, therefore the quantity
\beq\label{eq:magnitude}
U_{\alpha\beta}\sum_\bd{k}|\psi_{m,\bd{k}+\bd{A},\alpha}|^2|\psi_{m,\bd{k}-\bd{A},\beta}|^2
\eeq
provides a measure of ordering at each $\bd{A}$. Equation~\eqref{eq:magnitude} shares the spirit of the concepts of ``quantum geometric nesting" or ``discrepancy" in recent studies~\cite{han2024qgn,sun2025flat,zhang2026identifying}, but is stated for the half-filling, zero-temperature limit.

Similar to the analysis of condensation energy in (i), the function $|\psi_{m,\bd{k}+\bd{A},\alpha}|^2|\psi_{m,\bd{k}-\bd{A},\beta}|^2$ with fixed $\bd{k}$ can change with $\bd{A}$ drastically. However, the orbital composition in a band, $|\psi_{m,\bd{k},\alpha}|^2$ typically varies smoothly with $\bd{k}$. This leads to the fact that the summation over $\bd{k}$ in Eq.~\eqref{eq:magnitude} makes it a smooth function of $\bd{A}$. This point is further illustrated using 2D lattice models in Appendix~\ref{app:lattice}.

\tit{Comparison with BCS superconductors in dispersive bands.} In the dispersive-band case, due to the strong dispersion and the consequent energetic penalty for momenta away from the Fermi level, the order parameter vanishes (depairs) at a vector potential much smaller than the reciprocal lattice constant (see the illustration in Fig.~\ref{fig:bulk}(c)). In the flat band case, there is no energetic penalty, and the orbital structure of the wavefunctions, i.e., quantum geometry, determines pairing via the form factors. Therefore, (ii) is equivalent to stating that there is no complete depairing in flat bands in the limit considered.\\

\tit{Combining (i) and (ii), we conclude that half-filled flat-band superconductors at low temperatures in general have a negative free energy at all $\bd{A}$. Depairing due to a finite vector potential $\bd{A}$ does not occur and lead to a superconducting-normal transition, provided that the form factor of Eq.~\eqref{eq:uab} remains finite at least at some $\bd{k}$-points in the Brillouin zone.}
\section{Lattice models}
\label{app:lattice}
We use two 2D lattice models to illustrate the global negativity of free energy in flat bands and justify the cosine model of free energy. The free energy densities $f_s(\bd{A})$ of these lattice models often contain higher-order harmonic terms, which, however, only change the physics quantitatively, as we explain in Appendix~\ref{app:fixedpoint}.
\subsection{Bipartite Lieb lattice}
\label{app:lieb}
\begin{figure}[t!]
\centering
\includegraphics[width=0.48\textwidth]{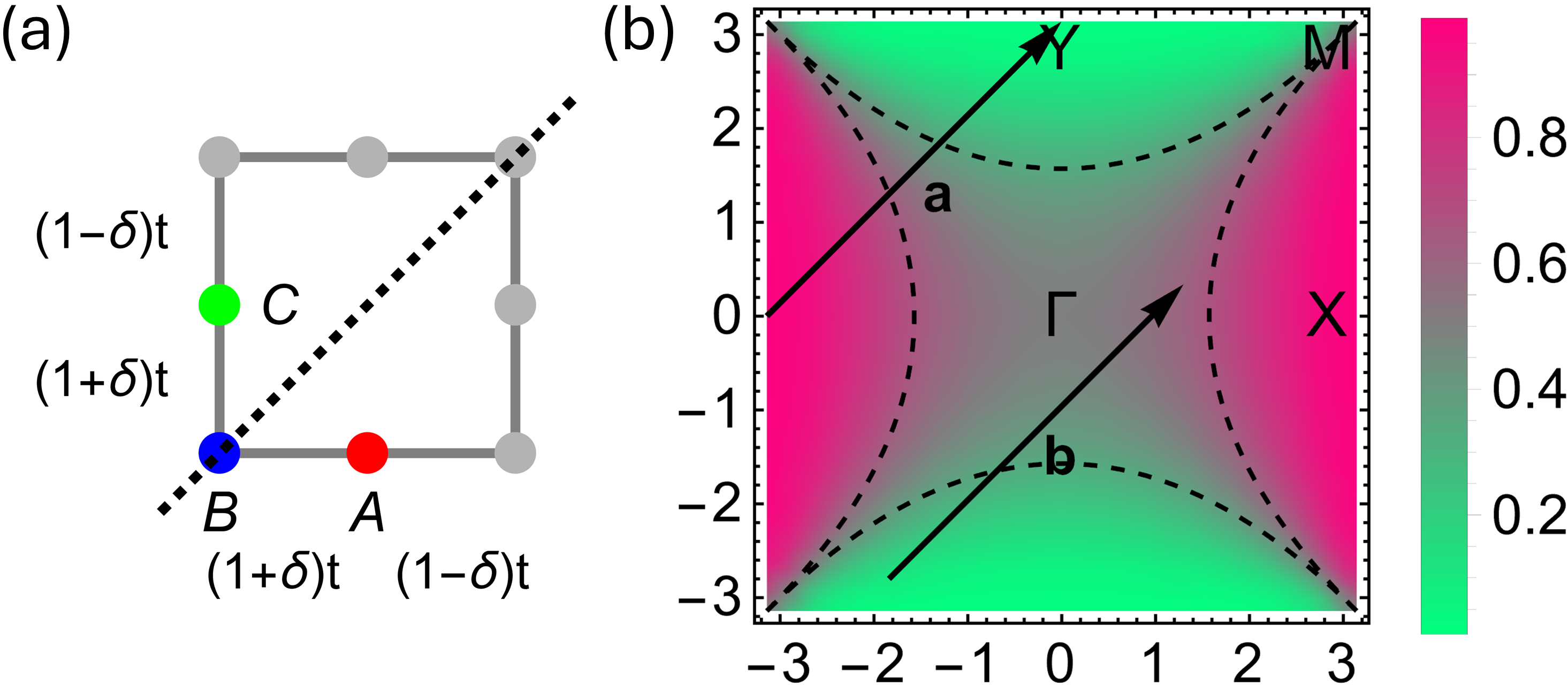}
\caption{(a) Unit cell of the Lieb lattice model with three orbitals. The dashed line indicates the mirror plane. (b) Density plot of the weight of orbital $A$, $|\psi_{0,\bd{k},A}|^2$ of the flat band in the Brillouin zone ($\bd{k}$-space), with the reddish (greenish) color corresponding to orbital $A$ ($C$), and $\Gamma,X,Y,M$ the four TRIM.}
\label{fig:lieb1}
\end{figure}

The tight-binding hopping graph of this model~\cite{julku2016geometric} is shown in Fig.~\ref{fig:lieb1}(a). It contains three orbitals ($A,B,C$) per unit cell and has staggered nearest-neighbor hoppings, $(1\pm\delta)t$. The non-interacting tight-binding Hamiltonian is
\beq\label{eq:hamlieb}
\hham_0=\sum_{\bd{k},\alpha\beta}c_{\bd{k}\alpha}^\dagger h_{\alpha\beta}(\bd{k})c_{\bd{k}\beta},
\eeq
where $c_\bd{k}^\dagger$ ($c_\bd{k}$) is the creation (annihilation) operator in the momentum basis. $\alpha,\beta$ run over $A,B,C$. The Bloch Hamiltonian
\beq\label{eq:hklieb}
h(\bd{k})=2t\begin{pmatrix}
0&a_\bd{k}^*&0\\
a_\bd{k}&0&b_\bd{k}\\
0&b_\bd{k}^*&0
\end{pmatrix}
\eeq
with
\begin{align}
&a_\bd{k}=\cos\frac{k_x}{2}+i\delta\sin\frac{k_x}{2}, \nonumber\\
&b_\bd{k}=\cos\frac{k_y}{2}+i\delta\sin\frac{k_y}{2}.
\end{align}
The spin indices are suppressed in $\hham_0$ above due to the absence of spin-orbital coupling. The energy spectrum of $h(\bd{k})$ consists of a flat band, $\varepsilon_{0,\bd{k}}=0$, which is separated from the upper and lower dispersive bands, $\varepsilon_{\pm,\bd{k}}$ by an energy gap $2\sqrt{2}t\delta$. The flat band has the normalized eigenstate
\beq
\psi_{0,\bd{k}}=\frac{1}{\sqrt{|a_\bd{k}|^2+|b_\bd{k}|^2}}\begin{pmatrix}
b_\bd{k}&0&-a_\bd{k}
\end{pmatrix}^T,
\eeq
which contains orbital $A$ and $C$ only, see Fig.~\ref{fig:lieb1}(b).

\begin{figure}[t!]
\centering
\includegraphics[width=0.48\textwidth]{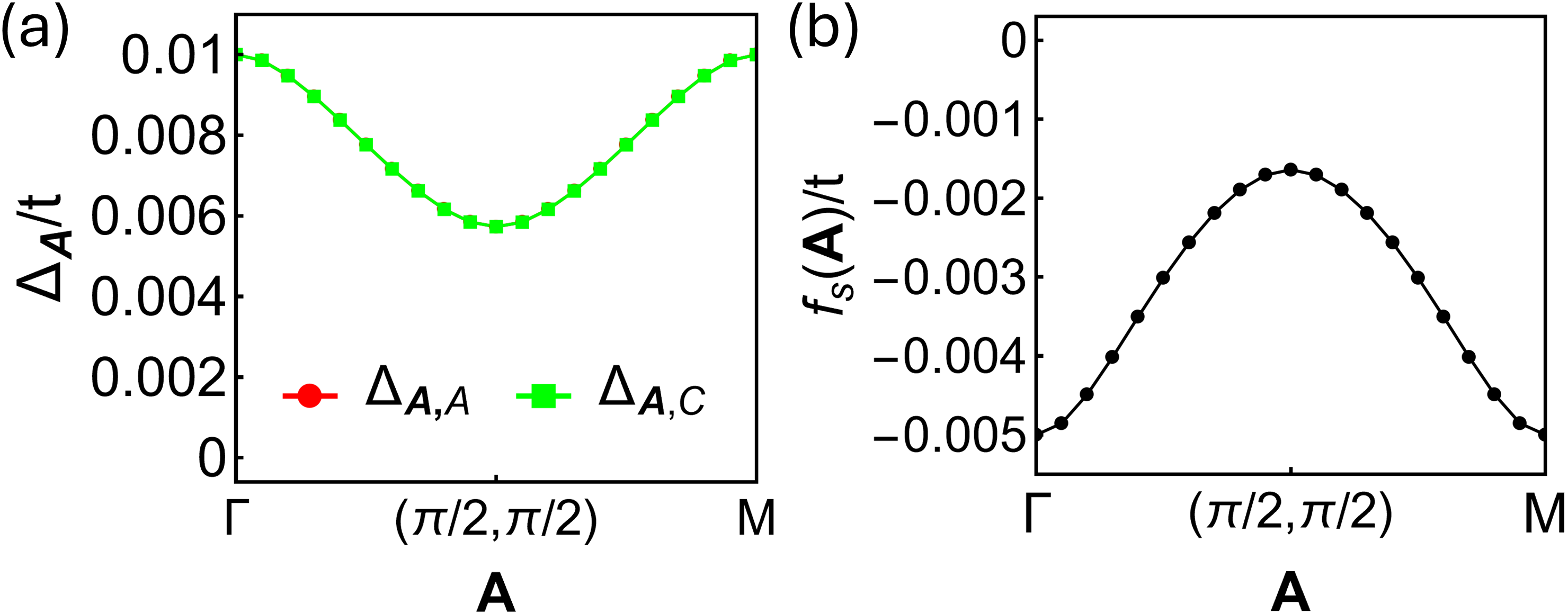}
\caption{Order parameters $\Delta_{\bd{A},\alpha}$ and free energy density $f_s(\bd{A})$ (in the units of $t$) along $\Gamma M$ in $\bd{A}$-space of the Lieb lattice model, with parameters $U=0.04t$, $\delta=0.1$. (a) The pairing potentials remain uniform ($\Delta_{\bd{A},A}=\Delta_{\bd{A},C}$) along $\Gamma M$, and do not completely depair. (b) $f_s(\bd{A})$ remains negative along $\Gamma M$.}
\label{fig:lieb2}
\end{figure}

The superconducting state is assumed to be concerned with the flat band only, as long as the superconducting gap $\Delta\ll2\sqrt{2}t\delta$ and its chemical potential $\mu\approx 0$. We consider an attractive Hubbard interaction,
\beq
\hham_I=-U\sum_i(\hat{n}_{iA\uparrow}\hat{n}_{iA\downarrow}+\hat{n}_{iC\uparrow}\hat{n}_{iC\downarrow}),
\eeq
Here, orbital $B$ is irrelevant as it is absent in the flat band. Then, mean-field decoupling leads to the following zero-temperature gap equation,
\beq\label{eq:gaplieb}
\Delta_{\bd{A},\alpha}=\frac{U}{2N_c}\sum_\bd{k}\psi_{\bd{k}-\bd{A},\alpha}^*\psi_{\bd{k}+\bd{A},\alpha}\frac{\Delta_{0,\bd{k}}(\bd{A})}{\sqrt{\mu_\bd{A}^2+|\Delta_{0,\bd{k}}(\bd{A})|^2}},
\eeq
where the order parameter in the orbital basis, $\Delta_{\bd{A},\alpha}$, and chemical potential $\mu_\bd{A}$ depend on $\bd{A}$ self-consistently (see Appendix~\ref{app:currentoperator} for details). The order parameter in the band basis is
\beq\label{eq:gapliebband}
\Delta_{0,\bd{k}}(\bd{A})=\sum_{\alpha=A,C}\psi_{0,\bd{k}+\bd{A},\alpha}^*\Delta_{\bd{A},\alpha}\psi_{0,\bd{k}-\bd{A},\alpha}.
\eeq
Equations~\eqref{eq:gaplieb}, \eqref{eq:gapliebband} are just Eq.~\eqref{eq:gapband} and~\eqref{eq:gap1} for intra-orbital pairings (with $\mu_\bd{A}=0$ for all $\bd{A}$ at half filling).

We now solve the gap equation and compute the free energy for the high-symmetry line $\Gamma(0,0)-M(\pi,\pi)$ in $\bd{A}$-space. The gap equation~\eqref{eq:gaplieb} is solved at half-filling using the Newton-Raphson iteration method. The mirror symmetry of the model about $(1,1)$ direction (see Fig.~\ref{fig:lieb1}(a)) justifies a real, uniform pairing solution to the gap equation, $\Delta_{\bd{A},A}=\Delta_{\bd{A},C}$ along $\Gamma M$, which are plotted in Fig.~\ref{fig:lieb2}(a). Then, the free energy density $f_s(\bd{A})=F_s(\bd{A})/N_c$ is calculated from Eq.~\eqref{eq:fa} (with band index $m=0$, and lattice constant $a=1$), which resembles the cosine function and is plotted in Fig.~\ref{fig:lieb2}(b).

To understand the dependence of $\Delta_{\bd{A}}$ on $\bd{A}$ in Fig.~\ref{fig:lieb2}(a), we follow Eq.~\eqref{eq:magnitude} and measure the ``degree of pairing" at $\bd{A}$ by
\beq\label{eq:magnitudelieb}
U\sum_\bd{k}|\psi_{0,\bd{k}+\bd{A},\alpha}|^2|\psi_{0,\bd{k}-\bd{A},\alpha}|^2
\eeq
(the result is the same for $\alpha=A$ and $C$ due to the symmetry between the two orbitals). The largest depairing refers to the minimum of Eq.~\eqref{eq:magnitudelieb}, which corresponds to a vector potential $\bd{A}$ that makes the orbital weight $|\psi_{0,\bd{k}+\bd{A},\alpha}|^2$ globally mismatch with $|\psi_{0,\bd{k}-\bd{A},\alpha}|^2$ in $\bd{k}$-space. This is best illustrated by the plot of the weight of orbital $A$ in the flat band, Fig.~\ref{fig:lieb1}(b), where the reddish (greenish) color means polarization to orbital $A$ ($C$). The band is fully polarized to orbital $A$ and $C$ at $\bd{k}=(-\pi,0)$ and $(0,\pi)$, respectively. Therefore, $2\bd{A}=(\pi,\pi)$ (arrows in Fig.~\ref{fig:lieb1}(b)) makes the orbital weight mismatch to be the largest for most of the $\bd{k}$ vectors in the Brillouin zone (in particular, examplified by arrow $\bd{a}$, which connects two $\bd{k}$-points fully polarized to $A$ and $C$, respectively). This explains why the largest depairing is at $\bd{A}=(\frac{\pi}{2},\frac{\pi}{2})$. A more mathematical treatment is provided in Ref.~\cite{zhang2026identifying} for general band geometry and pairing interactions.

However, a multi-orbital flat band can only change its orbital composition smoothly in $\bd{k}$-space. This leads to a large region of mixed orbital compositions, indicated by the star-shaped area enclosed by dashed curves in Fig.~\ref{fig:lieb1}(b). Owing to the presence of this region, there exist many arrows like $\bd{b}$ which connect $\bd{k}$-points that are not fully polarized to orbitals, leading to a finite contribution to Eq.~\eqref{eq:magnitudelieb}. As a result, the order parameters do not completely depair even at $\bd{A}=(\frac{\pi}{2},\frac{\pi}{2})$.

\subsection{Flattened Bernevig-Hughes-Zhang model}
\label{app:bhz}
The second model we choose is the two-band version of the Bernevig-Hughes-Zhang (BHZ) or Qi-Wu-Zhang model~\cite{bernevig2006quantum,qi2006topo} artificially flattened by long-range hoppings~\cite{jiang2023pdw}. With this model, we provide an example where the orbitals are less symmetric in the flat band than the Lieb lattice case.

\begin{figure}[t!]
\centering
\includegraphics[width=0.48\textwidth]{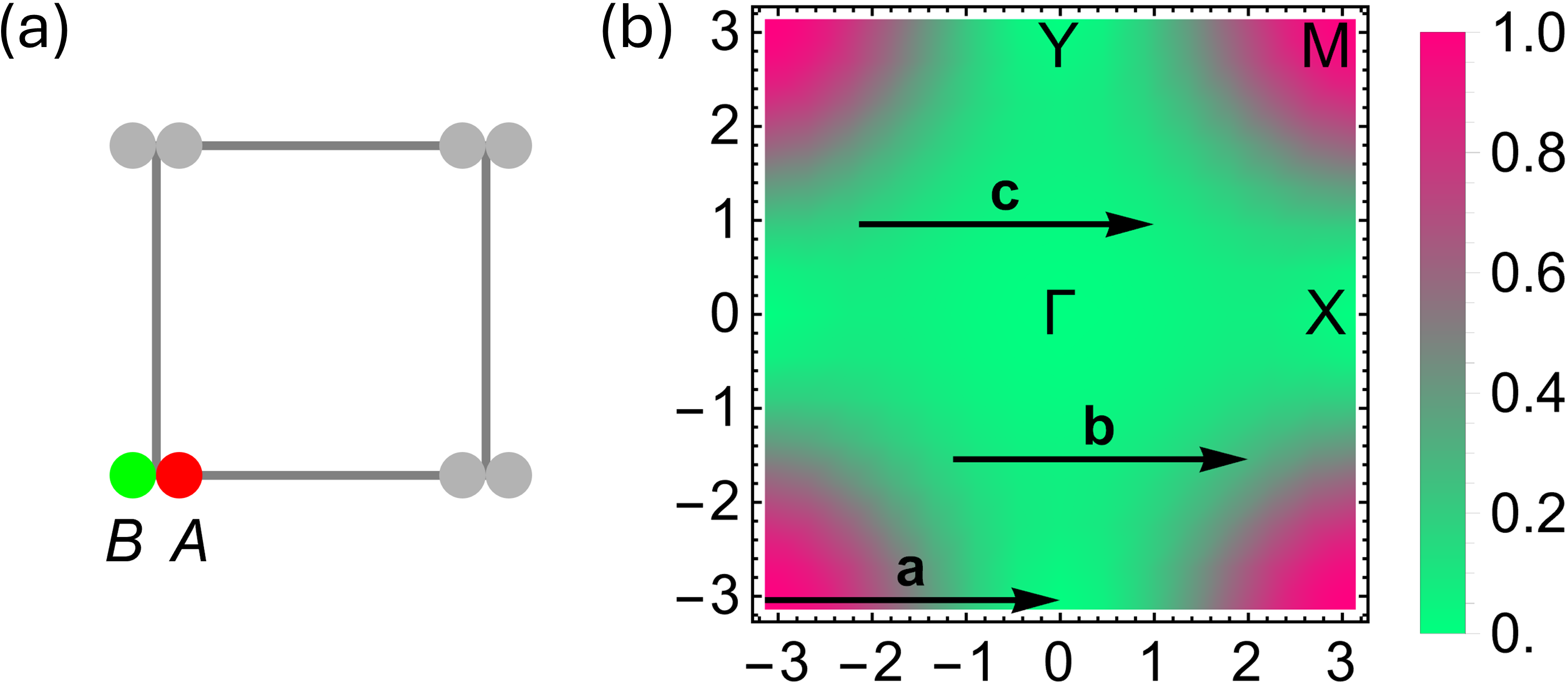}
\caption{(a) Unit cell of the square-lattice flattened BHZ model with two orbitals sitting at the same site. (b) Density plot of the weight of orbital $A$, $|\psi_{v,\bd{k},A}|^2$ of the valence band in the Brillouin zone ($\bd{k}$-space), with the reddish (greenish) color corresponding to orbital $A$ ($B$), and $\Gamma,X,Y,M$ the four TRIM.}
\label{fig:bhz1}
\end{figure}

\begin{figure}[t!]
\centering
\includegraphics[width=0.48\textwidth]{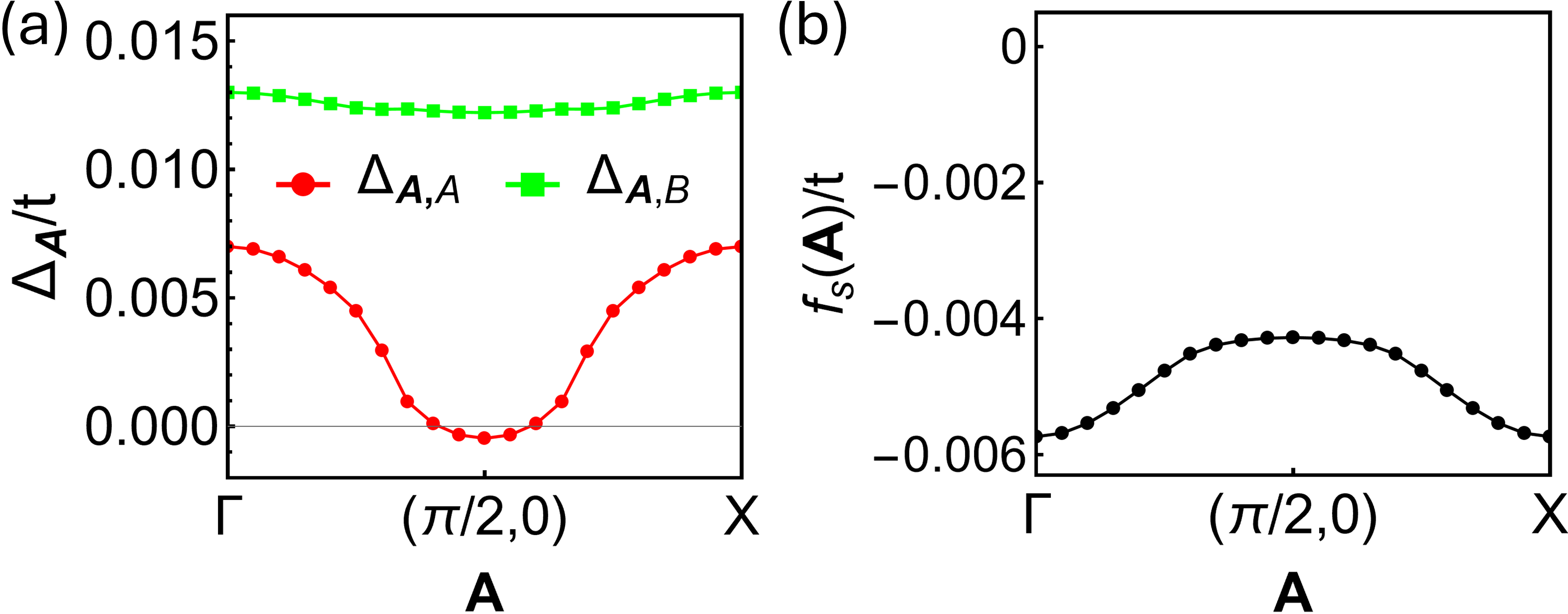}
\caption{(a) Order parameters $\Delta_{\bd{A},\alpha}$ and (b) free energy density $f_s(\bd{A})$ (in the units of $t$) along $\Gamma X$ in $\bd{A}$-space of the flattened BHZ model, with parameters $m=1$, $U_A=0.055t$, $U_B=0.035t$.}
\label{fig:bhz2}
\end{figure}

Fig.~\ref{fig:bhz1}(a) shows a unit cell of the flattened BHZ model, where the two orbitals ($A$ and $B$) sit at the same site. The tight-binding hopping graph is complicated in real space due to the band flattening. The Bloch Hamiltonian is
\beq\label{eq:hkbhz}
h(\bd{k})=t\frac{\bd{d}(\bd{k})}{|\bd{d}(\bd{k})|}\cdot\bds{\sigma},
\eeq
where the vector $\bd{d}(\bd{k})$ has components 
\begin{align}
&d_x(\bd{k})=\sin k_x,\nonumber\\
&d_y(\bd{k})=\sin k_y,\nonumber\\
&d_z(\bd{k})=m+\cos k_x+\cos k_y,
\end{align}
and $|\bd{d}(\bd{k})|$ is its magnitude. Here $\bds{\sigma}=(\sigma_x,\sigma_y,\sigma_z)$ are Pauli matrices acting on the two-orbital space. The Hamiltonian~\eqref{eq:hkbhz} has a spectrum of two flat bands, $\varepsilon_{\pm,\bd{k}}=\pm t$. We consider an attractive Hubbard interaction
\beq
\hham_I=-\sum_i(U_A\hat{n}_{iA\uparrow}\hat{n}_{iA\downarrow}+U_B\hat{n}_{iB\uparrow}\hat{n}_{iB\downarrow}),
\eeq
with $0<U_A,U_B\ll t$, and assume that the superconducting state is at the valence band.

Tuning the mass $m$ allows topological transitions between different phases of the model, and changes the orbital compositions of the two bands. Without loss of generality, we choose $m=1$, which corresponds to a topological phase with the valence band slightly polarized to orbital $B$ ($75\%$ $B$ and $25\%$ $A$, see Fig.~\ref{fig:bhz1}(b)). The BHZ model has $C_4$ rotation symmetry, so we solve the gap equation along $\Gamma(0,0)X(\pi,0)$ in $\bd{A}$-space using the same method as we use for the Lieb lattice model. In Fig.~\ref{fig:bhz2}, we plot the real solutions of the order parameters and the free energy density along $\Gamma X$ for chosen interactions, $U_A=0.055t$, $U_B=0.035t$.

Figure~\ref{fig:bhz2}(a) shows that $\Delta_{\bd{A},A}$ depairs strongly along $\Gamma X$; it completely depairs near $(\frac{\pi}{2},0)$, since $2\bd{A}=(\pi,0)$ (see arrow $\bd{a}$ in Fig.~\ref{fig:bhz1}(b)) makes an almost complete mismatch of its weight in $\bd{k}$-space. Conversely, $\Delta_{\bd{A},B}$ varies little with $\bd{A}$ due to the dominance of the weight of orbital $B$ in the band; even at $(\frac{\pi}{2},0)$, it is almost unaffected since many arrows like $\bd{b}$ and $\bd{c}$ exist that fail to connect two $\bd{k}$-points with mismatched weight of orbital $B$. Combining these two effects, $f_s(\bd{A})$ resembles a cosine function (Fig.~\ref{fig:bhz2}(b)). In this case, orbital $B$ forms a leading pairing background for the superconducting state provided that $U_B$ is attractive. We also note that although orbital $A$ takes up only $25\%$ in the band, it leads to a nonzero superfluid weight near the $\Gamma$-point as well as the oscillation for the cosine model.

\section{Solutions to the time-independent sine-Gordon equation}
\label{app:sge}
The kink and breather solutions to the time-independent sine-Gordon equation
\beq\label{eq:sgey}
y''(x)-a\sin(by(x))=0\,\,\,(a,b>0)
\eeq
can be obtained analytically, where for our case,
\beq
a=2\mu_0\delta_2\frac{ea}{\hbar},\,\,\,b=\frac{2ea}{\hbar}.
\eeq
Integrating Eq.~\eqref{eq:sgey} with $\derd y$ once leads to
\beq\label{eq:yprime}
y'=\pm\sqrt{-\frac{2a}{b}\cos(by)+c_1}
\eeq
where $c_1>0$ is a constant of integration. The $\pm$ sign labels the two branches of the solution, $y'>0$ or $<0$, corresponding to the charge of the soliton mode. For the solution to be real, we require $c_1\geq \frac{2a}{b}$.

To proceed, it is sufficient to restrict to the domain $0\leq y\leq \frac{2\pi}{b}$ for the $+$ case, or $-\frac{2\pi}{b}\leq y\leq0$ for the $-$ case, since the ``source" of Eq.~\eqref{eq:sgey}, $a\sin (by)$ has the periodicity of $\frac{2\pi}{b}$ and vanishes at $y=0$. Then, the solution being a kink or a breather depends on whether
\beq
y'|_{y=0}=0\,\,\,\text{or}\,\,\,\neq0,
\eeq
i.e., $c_1=\frac{2a}{b}$ or $c_1>\frac{2a}{b}$.
\subsection{Kink solution}
For the kink solution ($c_1=\frac{2a}{b}$), integration of Eq.~\eqref{eq:yprime} leads to
\beq
y=\pm\bigg[\frac{2}{b}\sin^{-1}\tanh(\sqrt{ab}(x-x_0))+\frac{\pi}{b}\bigg],
\eeq
where $x_0$ is the second constant of integration, corresponding to the center of the kink (can be taken as 0).
\subsection{Breather solution}
For the breather solution, since $c_1>\frac{2a}{b}$, we introduce another parameter $\alpha$ to replace $c_1$,
\beq
\frac{1}{\alpha}=\frac{1}{2}+\frac{c_1b}{4a},
\eeq
with $0<\alpha<1$. Notice that (we substituted $A_y\rightarrow y$)
\beq
y'|_{y=0}=\pm\sqrt{c_1-\frac{2a}{b}}\equiv B_\text{min},
\eeq
so $B_\text{min}$ and $\alpha$ has the relation
\beq
\frac{1}{\alpha}=1+\frac{b}{4a}B_\text{min}^2,
\eeq
which gives Eq.~\eqref{eq:1overalpha}.

Substituting the variable $y\rightarrow\frac{2\theta+\pi}{b}$ in Eq.~\eqref{eq:yprime}, we obtain
\beq\label{eq:elliptic}
\begin{split}
&x=\pm\sqrt{\frac{1}{ab}}\int\frac{\derd\theta}{\sqrt{\frac{1}{2}+\frac{c_1b}{4a}-\sin^2\theta}}\\
\Longrightarrow\,\,\,&x-x_0=\pm\sqrt{\frac{\alpha}{ab}}F(\theta|\alpha),
\end{split}
\eeq
where $F(\theta|\alpha)$ is the elliptic function of the first kind
\beq
F(\theta|\alpha)=\int_0^\theta\frac{\derd\theta'}{\sqrt{1-\alpha\sin^2\theta'}}.
\eeq
The inverse of Eq.~\eqref{eq:elliptic} is
\beq
\theta=\pm\text{am}\left(\sqrt{\frac{ab}{\alpha}}(x-x_0)\bigg|\alpha\right),
\eeq
which finally leads to Eq.~\eqref{eq:breather}.
\section{Quantized flux in magnetic field walls}
\label{app:fluxquantization}
In this section, we explain how the magnetic field walls manage to avoid requiring a normal core, focusing for simplicity on the kink solution. The central insight of this derivation is the importance of the lattice periodicity: at the points in the field wall where current vanishes, $A_y$ is quantized to multiples of $\frac{\pi\hbar}{ea}$, which yields a multiple of the flux quantum $\phi_0=\frac{\pi\hbar}{e}$ when integrated between lattice sites.

We begin by briefly reviewing the origin of flux quantization in BCS theory, where (per Eq.~\eqref{eq:BCSjdef})
\beq
\bd{j}=-D_s\bd{A}=-D_s\left(\bd{A}^{(\chi)}+\frac{\hbar}{2e}\nabla\varphi\right).
\eeq
Integrating over a closed loop $\gamma$ along which $\varphi$ is always defined,
\beq
\oint_\gamma\bd{j}\cdot \derd\bd{l}=-D_s\left(\oint_\gamma\bd{A}^{(\chi)}\cdot \derd\bd{l}+\frac{\hbar}{2e}\oint_\gamma\nabla\varphi\cdot \derd\bd{l}\right).
\eeq
For simplicity, we consider only loops along which $\bd{j}\cdot \derd\bd{l}=0$. The magnetic flux $\phi$ then quantizes as
\beq\label{eq:quantcond}
\phi=\oint_\gamma\bd{A}^{(\chi)}\cdot \derd\bd{l}=-\frac{\hbar}{2e}\oint_\gamma\nabla\varphi\cdot \derd\bd{l}=N\phi_0
\eeq
where $N$ is an integer.

\begin{figure}
\centering
\includegraphics[width=0.75\linewidth]{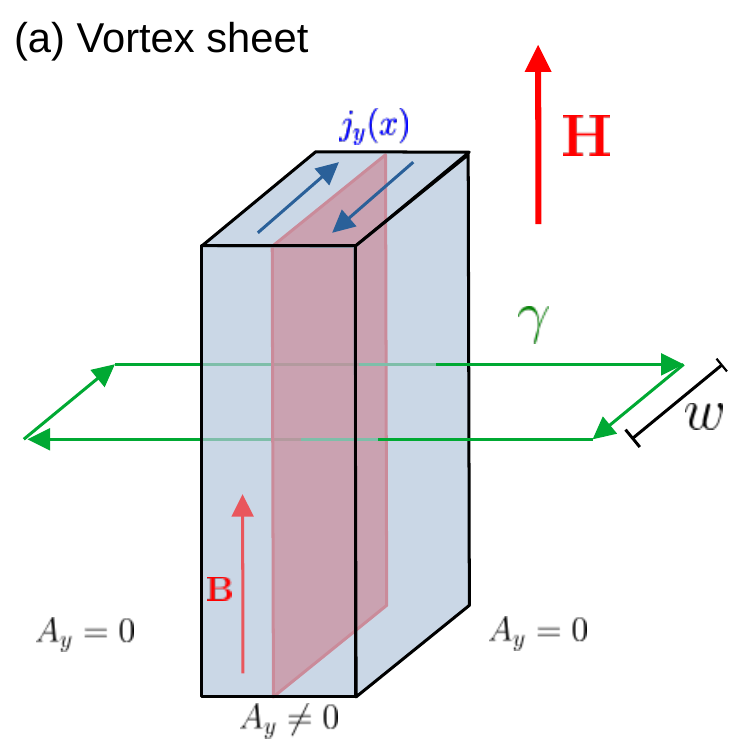}
\includegraphics[width=0.75\linewidth]{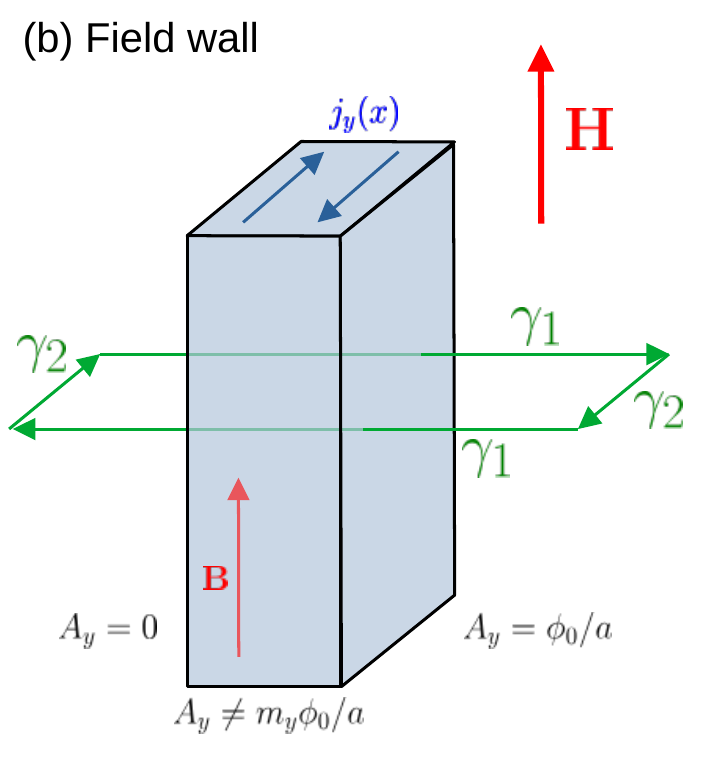}
\caption{Both field walls and vortex sheets feature counterpropagating currents, but they are topologically distinct objects. (a) In vortex sheets, the gauge-invariant vector potential $\bd{A}$ vanishes far from the sheet. Because of this, there is necessarily a normal core (red plane), which evades flux quantization around the green loop $\gamma$. (b) In magnetic field walls, values of $\bd{A}$ instead correspond to distinct TRIM points on the two sides of the field wall. There is no normal core, so flux quantization is maintained: if each segment of $\gamma_2$ traverses a lattice vector, then the integral of $\bd{A}$ is a multiple of the flux quantum. The geometry has been chosen so that the vector potential $\bd{A}$ can have non-zero components only in the $y$ direction.}
\label{fig:fluxcontour}
\end{figure}

The quantization condition Eq.~\eqref{eq:quantcond} is generally heavily constraining for translation-invariant 2D defects such as vortex sheets \cite{LandauLifshitz1955,volovik2003universe}. Integrating along the contour in Fig.~\ref{fig:fluxcontour}(a), the integrated flux can be changed continuously by changing the width $w$ of the rectangular region. The resulting continuously varying flux would violate the quantization condition Eq.~\eqref{eq:quantcond}, which was based on having a continuous and well-defined condensate phase $\phi$ at all locations. To resolve the contradiction (while keeping $\bd{A}^{(\chi)}$ smooth), there must be a discontinuous jump in $\phi$ somewhere along the contour, indicating that there must be a normal core. This normal core is avoidable with certain specific, complicated order parameters, such as that of Helium-3 \cite{volovik2003universe}, but otherwise it is a topological requirement.

In the cosine model, the above characterization breaks down because it is no longer the case that $\bd{j}=-D_s\bd{A}$; this must be replaced by the relationship between current and free energy given in Eq.~\eqref{eq:jdef}. Nevertheless, as we show below, the flux quantization argument can be generalized to the magnetic field walls, without using the relation $\bd{j}=-D_s\bd{A}$.

In the region far from the field walls, $\bd{A}$ is approximately an equilibrium solution $\bd{A}_\bd{m}=\frac{\phi_0}{a}\bd{m}$ with $\bd{m}=(m_x,m_y)$ a vector of integers (in the geometry of Fig.~\ref{fig:fluxcontour}(b) $m_x=0$). Note that
\beq
\int_{\bd{r}_0}^{\bd{r}_0+a(n_x,n_y)}\bd{A}_\bd{m}\cdot \derd\bd{l}=\phi_0(n_xm_x+n_ym_y)=\tilde N\phi_0
\eeq
is an integer multiple of the flux quantum. Hence, when the equilibrium solution is integrated from one site to another, the result is quantized if the two sites have the same sublattice index.

In the BCS case, we took the loop $\gamma$ to have $\bd{j}\cdot \derd\bd{l}=0$ along its length; for the cosine model, the argument is more subtle. We consider a loop $\gamma$ which can be decomposed into collections of segments $\gamma_{1,2}$. We demand that the segments in $\gamma_1$ be such that $\bd{A}\cdot d\bd{l}=0$ along their length, and that the segments in $\gamma_2$ be such that $\bd{A}=\bd{A}_\bd{m}$ along their length (potentially a different $\bd{m}$ for each curve segment in $\gamma_2$). Note that the curve in Fig.~\ref{fig:fluxcontour}(b) has this property: $\gamma_1$ consists of the segments passing through the field wall, and $\gamma_2$ consists of the segments far from the field wall.

In this case, the flux quantization equation \eqref{eq:quantcond} is modified to
\begin{align}
\phi&=\oint_\gamma\bd{A}^{(\chi)}\cdot \derd\bd{l}=\oint_{\gamma} \bd{A}\cdot \derd\bd{l} - \frac{\hbar}{2e}\oint_\gamma\nabla\varphi\cdot \derd\bd{l} \nonumber \\
&=\int_{\gamma_2} \bd{A}_\bd{m}\cdot \derd\bd{l}=\tilde N\phi_0,
\end{align}
showing that the magnetic flux is still quantized around such loops $\gamma$. (Note $\oint\nabla\varphi\cdot d\bd{l}$ vanishes because $\varphi$ is globally defined in the magnetic field wall, which is a simply-connected superconducting region.) More macroscopically, the magnetic flux \textit{per length of unit cell} is quantized to multiples of $\phi_0/a$. This quantization follows from the topological charge of the kink soliton (ultimately rooted in the periodicity of $f_s(\bd{A})$).
%
\section{Calculation of the lower critical field}
\label{app:hc1}
The lower critical field is calculated by equating the free energy $F_s^H$ of the superconducting state with a single kink to that of the zero-field state:
\beq
\int\derd^3r[f_s^H(\bd{r},\text{kink})-f_s^H(\bd{r},\text{zero-field})]=0.
\eeq
Since both states are translationally invariant in the $yz$ plane, it can be expressed as an integral along the $x$ direction,
\beq\label{eq:xint}
S_{yz}\int_{-\infty}^\infty\derd x[f_s^H(x,\text{kink})-f_s^H(x,\text{zero-field})]=0,
\eeq
where $S_{yz}$ is the area of kink wall in the $yz$ plane. Here
\begin{align}
&f^H_s(x,\text{kink})=f_s(A_y(x))+\frac{1}{2\mu_0}B_z(x)^2-HB_z(x) \nonumber\\
=&-\delta_1-\delta_2\cos\bigg(\frac{2ea}{\hbar}A_y(x)\bigg)+\frac{1}{2\mu_0}B_z(x)^2-HB_z(x),
\end{align}
with $A_y(x)$ the kink solution given by Eq.~\eqref{eq:kink} (consider the charge $+1$ case), and $B_z(x)$ its derivative,
\beq
B_z(x)=A_y'(x)=\frac{\hbar}{ea\lambda_L}\text{sech}\left(\frac{x}{\lambda_L}\right).
\eeq
Furthermore,
\beq
f^H_s(x,\text{zero-field})=f_s(0)=-\delta_1-\delta_2.
\eeq

It is convenient to introduce dimensionless variables
\begin{align}
&\tilde{A}_y=A_y/\bigg(\frac{\hbar}{ea}\bigg), \nonumber\\
&\tilde{x}=\frac{x}{\lambda_L}, \nonumber\\
&\tilde{H}=H/\bigg(\frac{\hbar}{\mu_0ea\lambda_L}\bigg), \nonumber\\
&\tilde{B}_z=B_z/\bigg(\frac{\hbar}{ea\lambda_L}\bigg).
\end{align}
Then the equation to solve becomes (besides a factor $\delta_2\lambda_LS_{yz}$)
\beq
\int_{-\infty}^{\infty}\derd\tilde{x}\big[1-\cos(2\tilde{A}_y(\tilde{x}))+2\tilde{B}_z(\tilde{x})^2-4\tilde{H}_{c1,w}\tilde{B}_z(\tilde{x})\big]=0.
\eeq
Performing the integral gives $\tilde{H}_{c1,w}=2/\pi$.
\section{Maximum and minimum $B$ fields}
\label{app:evolution}
Using the relations
\begin{align}
&A_y'(x)=B_z(x),\nonumber\\
&B_z'(x)=-\mu_0j_y(x),\nonumber\\
&j_y(x)=-\frac{\derd}{\derd A_y}f_s(A_y(x)),
\end{align}
we find
\beq\label{eq:increase}
\frac{1}{2\mu_0}\int \derd (B_z(x))^2=\int j_y(x)\derd A_y=-\int\derd f_s.
\eeq
Integrating from $B_\text{min}$ to $B_\text{max}$, this equation leads to
\beq\label{eq:conserv}
\frac{1}{2\mu_0}(B_\text{max}^2-B_\text{min}^2)=f_{s,\text{max}}-f_{s,\text{min}}=2\delta_2,
\eeq
which holds for the breather mode of any $\alpha$ value (including $\alpha\rightarrow1$, the kink mode). This relation dictates that as $B_\text{min}$ increases (consider the charge $+1$ case, $B_\text{min}>0$) or $\alpha$ decreases, $B_\text{max}$ also increases, and finally the two become almost equal. Using the relation between $B_\text{min}$ and $\alpha$, we obtain Eq.~\eqref{eq:almostequal}. Their difference vanishes as
\beq
B_\text{max}-B_\text{min}=\frac{2\mu_0(f_{s,\text{max}}-f_{s,\text{min}})}{B_\text{max}+B_\text{min}}\approx \sqrt{\alpha\mu_0\delta_2}.
\eeq
\section{Dynamics of the Maxwell equations}
\label{app:fixedpoint}
\tit{Summary}: Equations~\eqref{eq:jdef} and~\eqref{eq:maxwell} can be viewed as a periodic dynamical system governing the flow of the phase point $(\bd{A},\bd{B})$ with spatial coordinates. Properties of the point $(\bd{0},\bd{0})$, i.e., the zero-field state, in the phase space are crucial, since the kink modes arise from disturbances from it at $H_{c1,w}$. We find that $(\bd{0},\bd{0})$ is an attracting fixed point of the dynamical system~\cite{strogatz}. Consequently, a streamline starting near $(\bd{0},\bd{0})$ either flows to it or to its replica shifted by TRIM vector potentials in the phase space, with the latter situation corresponding to the formation of kink walls.\\

Since the system is homogeneous along $z$ direction, we impose $B_x=B_y=0$, $A_z=0$, $\partial_zA_x=\partial_zA_y=0$. The TRIM periodicity in $\bd{A}$ implies that the phase space of $(\bd{A},\bd{B})$, i.e., $(A_x,A_y,B_z)$, is the product of a torus and a line, $S^1\times S^1\times R$, but we consider the extended Brillouin zone for $A_x$ and $A_y$ for convenience.

Equation~\eqref{eq:maxwell} contains three equations
\beq\label{eq:dynamic1}
\begin{split}
&\partial_xA_y-\partial_yA_x=B_z,\\
&\partial_yB_z=\mu_0j_x(A_x,A_y),\\
&\partial_xB_z=-\mu_0j_y(A_x,A_y),
\end{split}
\eeq
for three variables $A_x$, $A_y$, $B_z$, and involves four velocity components, $\partial_xA_y$, $\partial_yA_x$, $\partial_xB_z$, $\partial_yB_z$. To be a fixed point requires the velocities to vanish, leading to:
\beq
B_z=0,\,\,\,j_x(\bd{A})=j_y(\bd{A})=0.
\eeq
Since we assume that $\bd{A}=0$ is the global minimum of $f_s(\bd{A})$, $\bd{j}(\bd{A})=-\nabla_\bd{A}f_s(\bd{A})$ vanishes there and $(\bd{0},\bd{0})$ is a fixed point.

Next, we show that $(\bd{0},\bd{0})$ is attracting in the periodic dynamical system along the high-symmetry line directions. A high-symmetry line (of $f_s(\bd{A})$ in $\bd{A}$-space) forces the current density $\bd{j}$ to be parallel to the line direction as $\bd{A}$ moves along the line. Without loss of generality, we assume the high-symmetry line is along the $y$ axis, so the relevant field components are $A_y$, $B_z$ and $j_y$. These components must be functions of $x$ only; if otherwise, they have $y$-dependence, e.g., $B_z=B_z(x,y)$, then $j_x=\frac{1}{\mu_0}\partial_y B_z\neq 0$, contradicting $\bd{j}\parallel\hat{\bd{y}}$. Therefore, Eq.~\eqref{eq:dynamic1} reduces to
\beq\label{eq:dynamic2}
\begin{split}
&\partial_xA_y(x)=B_z(x),\\
&\partial_xB_z(x)=-\mu_0j_y(A_y).
\end{split}
\eeq

\begin{figure}[t!]
\centering
\includegraphics[width=0.37\textwidth]{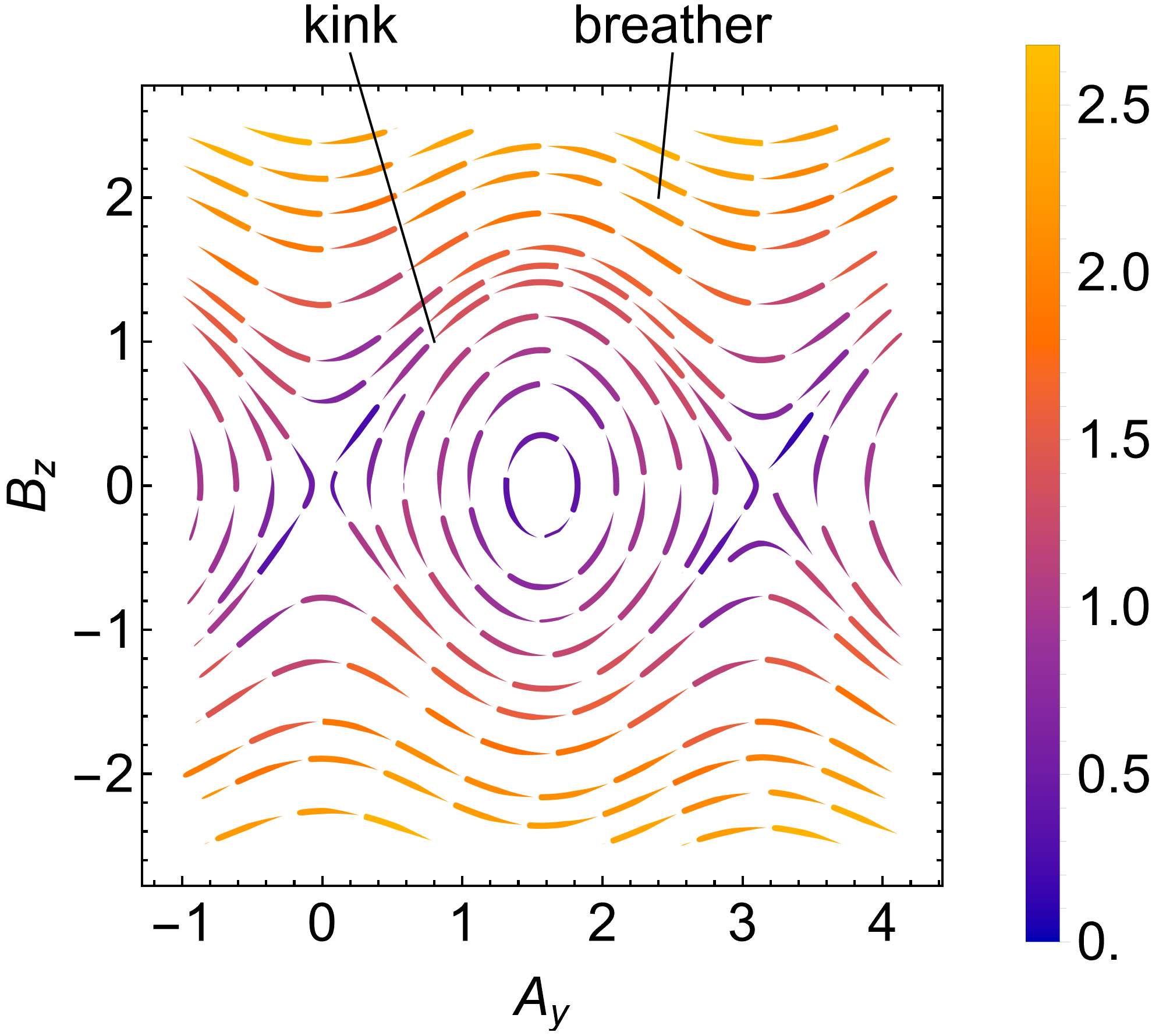}
\caption{Velocity field $(B_z,\sin(2A_y))$ and streamlines in the phase space of $(A_y,B_z)$ for the cosine model. Colors show the magnitude of the velocity, $\sqrt{B_z^2+\sin^2(2A_y)}$.}
\label{fig:fields}
\end{figure}

We now consider the current density of the cosine model,
\beq
j_y(A_y)=-2\delta_2\frac{ea}{\hbar}\sin\left(\frac{2ea}{\hbar}A_y\right),
\eeq
and show that the phase point $(A_y=0,B_z=0)$ is simultaneously a sink and source of the dynamical system, Eq.~\eqref{eq:dynamic2}. To illustrate this, we plot the velocity fields $(A_y'(x),B_z'(x))=(B_z,\sin(2A_y))$ (the constants $\mu_0$, $e$, $a$, $\hbar$, $\delta_2$ are taken to be 1) in the phase space of $(A_y,B_z)$ in Fig.~\ref{fig:fields}, which shows that the streamlines can either come into or leave from $(0,0)$. Point $(A_y=\frac{\pi}{2},B_z=0)$ is also a fixed point, but is a whirlpool center (neither a sink nor a source). Note: $A_y=\frac{\pi}{2}$ is the local maximum of the cosine model, which is energetically unfavorable.

The streamlines of the kink and the breather solitons are indicated in Fig.~\ref{fig:fields}. The kink mode has a relatively small $B_z$ coordinate, and its streamline is essentially constrained in $n\pi\leq A_y\leq (n+1)\pi$, with $n$ an integer, since the velocity vanishes at $(n\pi,0)$ (indicated by colors). Therefore, the streamline starting from $(0,0)$ must end at a TRIM, which is, itself, on the torus. Such a property about a fixed point is called ``attracting but not Lyapunov-stable"~\cite{strogatz}. On the contrary, the breather mode has a large $B_z$ coordinate, which instead can flow in the entire domain $-\infty\leq A_y\leq\infty$.

Below, we explain how a kink wall is related to the high-symmetry line. Imagine a phase point perturbed from $(0,0)$ towards the high-symmetry line direction, i.e., its $A_y$ becomes nonzero, and we study its subsequent flow. Using Eq.~\eqref{eq:dynamic2}, we find that Eq.~\eqref{eq:increase} holds, which leads to an indefinite integral in the phase space,
\begin{align}
&\frac{1}{2\mu_0}\int_{(0,0)}\derd B_z^2=\int_{(0,0)}\derd f_s \nonumber\\
\Longrightarrow\,\,\,&\frac{1}{2\mu_0}B^2_{z,\text{end}}=f_{s,\text{end}}-f_{s,(0,0)}.
\end{align}
Since the streamline can only end at a fixed point, where $B_z=0$, then we have $f_{s,\text{end}}=f_{s,(0,0)}$. In the case when TRIMs are the only global minima of $f_s$ in the $\bd{A}$-space, the high-symmetry line must connect two TRIMs, which means that both sides of the wall correspond to the zero-field state. If along the high-symmetry line there is another global minimum of $f_s$ with energy equal to the $\Gamma$-point, then the streamline ends there, which means the other side of the wall is a zero-field, but finite-momentum pairing state. We note that the analysis above relies on the periodicity of the dynamic system rather than the details of the cosine model; including high-order harmonic terms into $f_s(\bd{A})$ changes the results only quantitatively.

\section{Energy of a wall array vs. a vortex lattice}
\label{app:wallarray}
We compare the energy of a sparse wall array (see Fig.~\ref{fig:bulk}(c)) with a vortex lattice of equal density, i.e., the separation between neighboring kinks, $l$, in the array is equal to the distance between neighboring vortices in a vortex lattice. When these magnetic flux textures are stabler than the zero-field state, the last term of Eq.~\eqref{eq:fsh}, $-\bd{H}\cdot\bd{B}(\bd{r})$ dominates over the other terms; therefore, the total energy gain at fixed $H$ is approximately
\beq\label{eq:lz}
L_z\int \derd^2 r \bd{H}\cdot\bd{B}(\bd{r})=L_zH\Phi,
\eeq
where $L_z$ is the sample dimension along $z$ direction, and $\Phi=\int \derd^2 r B_z(\bd{r})$ is the total flux through a magnetic field texture.

The average magnitude of the $\bd{B}$ field in the flux region is $\sim\phi_0/a\lambda_L$ for walls vs. $\sim\phi_0/\lambda_L^2$ for vortices, while the area where the flux penetrates through is $\sim\lambda_Ll$ for walls vs. $\sim\lambda_L^2$ for vortices. Therefore, the ratio between the total fluxes in the two magnetic field textures is
\beq
\frac{\Phi_w}{\Phi_v}\sim\frac{\frac{\phi_0}{a\lambda_L}\cdot\lambda_Ll}{\frac{\phi_0}{\lambda_L^2}\cdot\lambda_L^2}=\frac{l}{a}\gg 1.
\eeq
We conclude that the wall array absorbs much more magnetic flux than the vortex lattice, therefore, according to Eq.~\eqref{eq:lz}, the former has a much lower energy than the latter. For the energy of a vortex lattice to be comparable to a wall array, the vortex lattice has to be much denser than the wall array, requiring $H_{c1,v}<H_{c1,w}$.

\section{Vortex of the cosine model in the extreme type-II limit}
\label{app:vortex}
We calculate the lower critical field, $H_{c1,v}$, of the cosine model for the vortex state in the extreme type-II limit, i.e., when the vortex core has a radius $r_0\ll\lambda_L$. In the process, we compare the behavior of $H_{c1,v}$ between the flat and the dispersive bands in this limit.

When $\kappa=\lambda_L/\xi\gg1$ ($\xi$ is typically of the same size as the core radius $r_0$), we still use the local self-consistency approximation and solve Eq.~\eqref{eq:maxwell} in cylindrical coordinates~\cite{campbell1972}. Note that in multi-orbital systems, there can be multiple $\lambda_L$ and $\xi$, and they can be anisotropic; by $\lambda_L\gg\xi$, we mean that the variations of order parameters in real space are much slower than that of the magnetic field, such that the nonlocal effect can be neglected.

We generalize the cosine model, Eq.~\eqref{eq:cosine}, to circular geometry,
\beq\label{eq:circular}
f_s(\bd{A})=-\delta_1-\delta_2\cos \left(\frac{2ea}{\hbar}|A|\right),
\eeq
where $|A|$ is the magnitude of $\bd{A}$. The model of Eq.~\eqref{eq:circular} takes into account the boundedness of $f_s(\bd{A})$ and neglects the lattice anisotropy. Then, we write $\bd{A}(\bd{r})=A(r)\hat{\bd{e}}_\varphi$, $\bd{B}(\bd{r})=B(r)\hat{\bd{z}}$, and $\bd{j}(\bd{r})=j(r)\hat{\bd{e}}_\varphi$, and Eq.~\eqref{eq:maxwell} becomes
\beq\label{eq:abcylinder}
A'(r)+\frac{A(r)}{r}=B(r),\,\,\,-B'(r)=\mu_0j(r).
\eeq
By the definition of current density, Eq.~\eqref{eq:jdef}, we have
\beq
\bd{j}=-\frac{2ea}{\hbar}\delta_2\sin\left(\frac{2ea}{\hbar}|A|\right)\hat{\bd{e}}_\bd{A}=-\frac{2ea}{\hbar}\delta_2\sin\left(\frac{2ea}{\hbar}A\right)\hat{\bd{e}}_\varphi ,
\eeq
where $\hat{\bd{e}}_\bd{A}$ is the unit vector along the direction of $\bd{A}$; $A(r)<0$, so $\hat{\bd{e}}_\bd{A}=-\hat{\bd{e}}_\varphi$. This leads to
\beq
j(r)=-\frac{2ea}{\hbar}\delta_2\sin\left(\frac{2ea}{\hbar}A(r)\right).
\eeq
We introduce dimensionless quantities
\beq\label{eq:dimensionless}
\begin{split}
&\tilde{r}=r/\lambda_L,\\
&\tilde{A}=A/(\hbar/ea),\\
&\tilde{B}=B/(\hbar/ea\lambda_L),\\
&\tilde{H}=H/(\hbar/\mu_0ea\lambda_L).
\end{split}
\eeq
Then, Eq.~\eqref{eq:abcylinder} leads to the following equation of $\tilde{A}(\tilde{r})$:
\beq\label{eq:radialcos}
\tilde{A}''+\frac{\tilde{A}'}{\tilde{r}}-\bigg[\frac{\tilde{A}}{\tilde{r}^2}+\frac{1}{2}\sin(2\tilde{A})\bigg]=0.
\eeq

\begin{figure}[t!]
\centering
\includegraphics[width=0.18\textwidth]{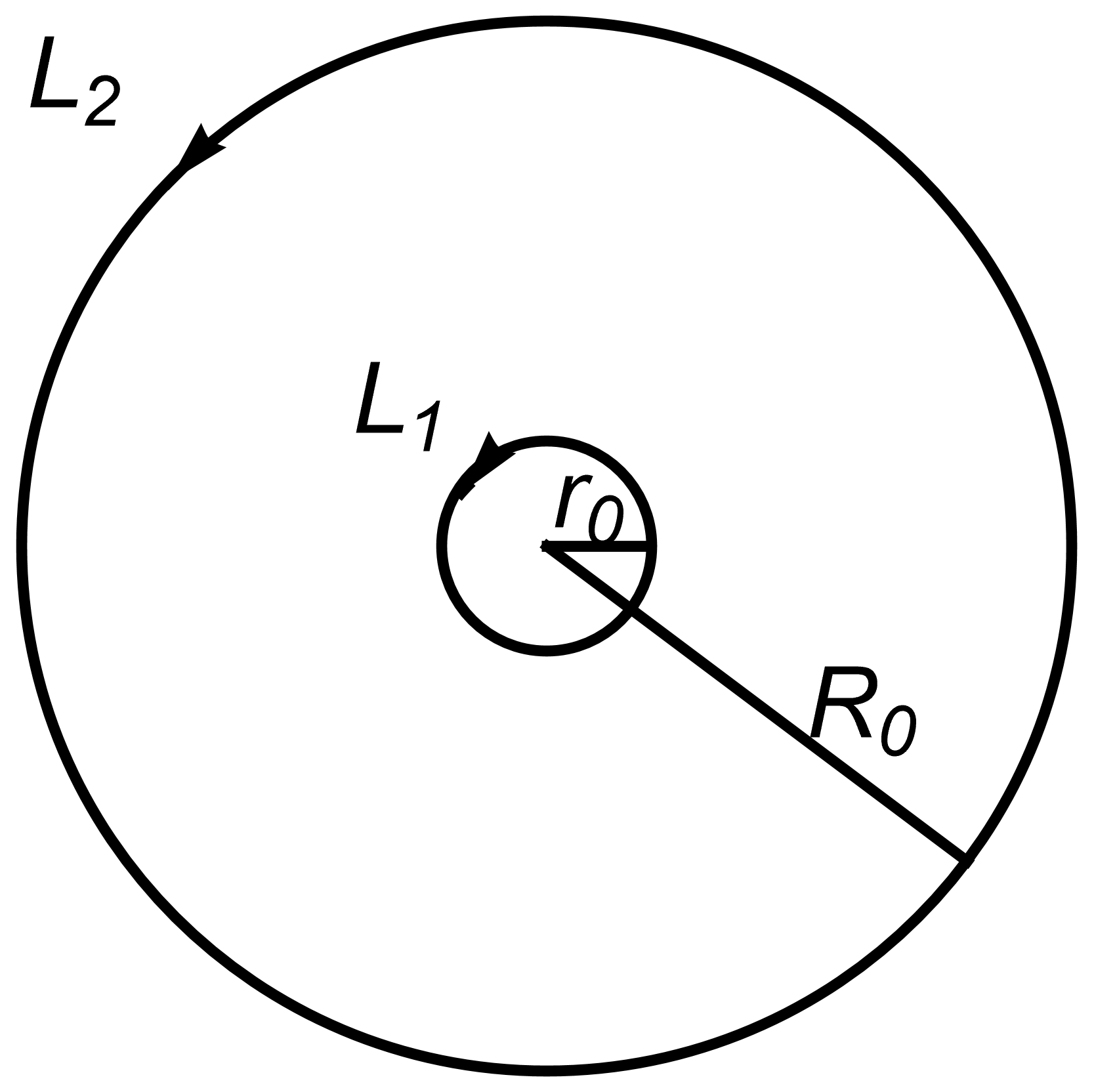}
\caption{A vortex split into the normal core region, $r<r_0$, and the superconducting outer region, $r_0<r<R_0$, with $R_0$ a few times larger than the penetration depth. Here $L_1$ and $L_2$ are two loops of radius $r_0$ and $R_0$, respectively, enclosing the center counterclockwise.}
\label{fig:vortex}
\end{figure}

Like the usual treatment of a single vortex in the $\kappa\gg1$ limit~\cite{abrikosov}, we split the vortex into the core region, $r<r_0$, and the outer region, $r_0<r<R_0$, where $R_0$ is equal to a few times $\lambda_L$ (see Fig.~\ref{fig:vortex}). The core is assumed to be completely normal, so $B(r)=\mu_0H$ there, and hence we neglect the contribution from the term of the gradient on the amplitude of the order parameter to the free energy.

The gauge-invariant vector potential $A(r)$ exponentially decays as $r\rightarrow\infty$, therefore, the outer loop integral at $R_0$ vanishes, $\oint_{L_2}\bd{A}\cdot\derd\bd{l}=0$. The flux quantization condition, which arises from the quantized accumulated phase change of the order parameter along a closed loop, is thus
\beq\label{eq:quantization}
\oint_{L_1}\bd{A}\cdot\derd\bd{l}=-\phi_0.
\eeq
Put into dimensionless quantities, Eq.~\eqref{eq:quantization} becomes
\beq\label{eq:quantization2}
\tilde{A}(\tilde{r}_0)=-\frac{\beta}{2\tilde{r}_0}.
\eeq
where $\beta\equiv a/\lambda_L$. For typical superconductors with dispersive bands, $\beta\sim10^{-3}-10^{-2}$; for flat bands, its value depends on the strength of interaction and quantum geometry of the band. Before solving Eq.~\eqref{eq:radialcos}, we derive the single vortex solution to the London model for dispersive bands.

\subsection{The London model of dispersive bands}
\begin{figure}[t!]
\centering
\includegraphics[width=0.48\textwidth]{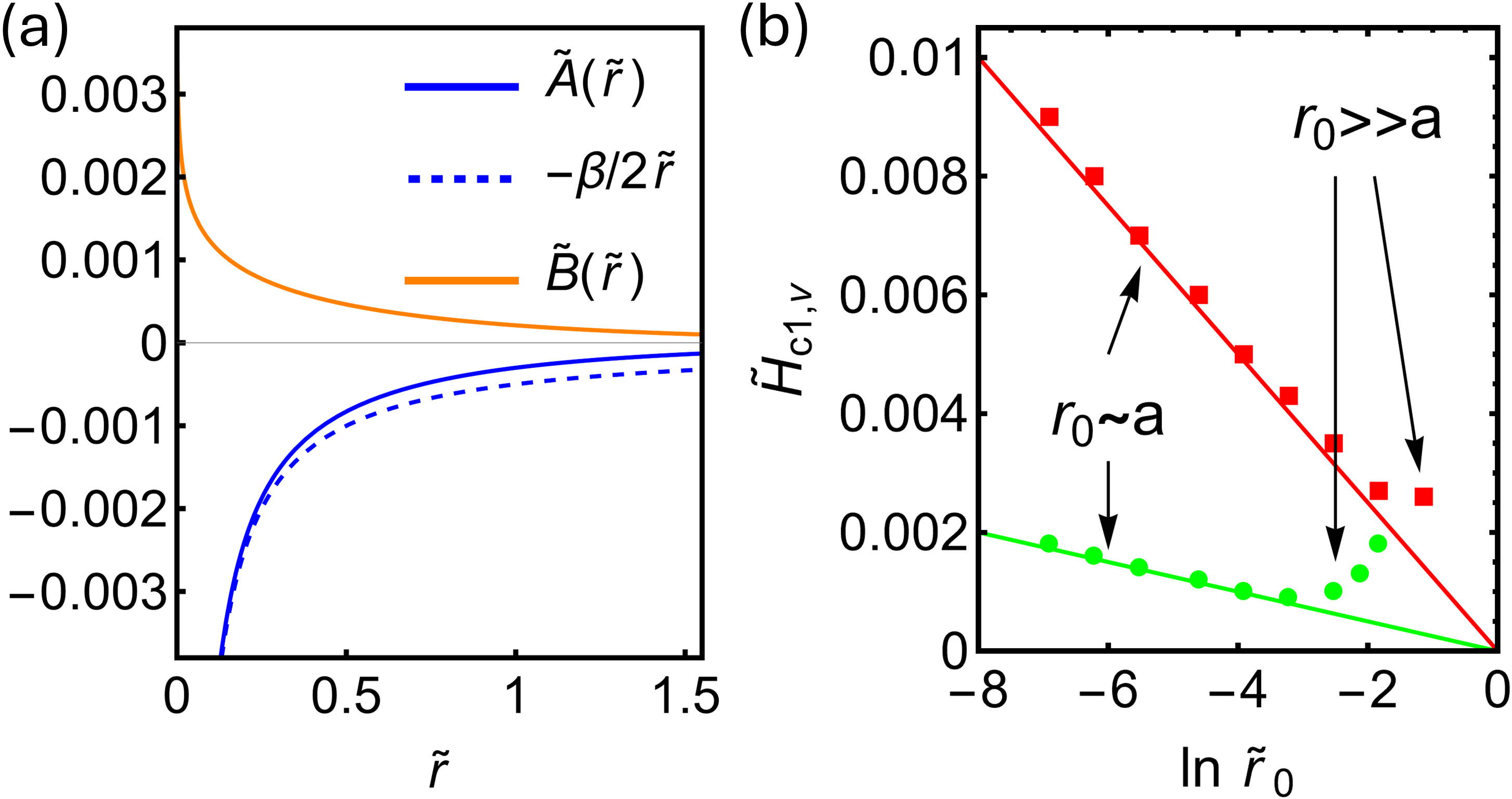}
\caption{(a) Comparison between functions $\tilde{A}(\tilde{r})$ (Eq.~\eqref{eq:bessel}), its associated $B$ field, $\tilde{B}(\tilde{r})$, and $-\beta/2\tilde{r}$, with $\beta=0.001$. (b) Lower critical field $\tilde{H}_{c1,v}$ vs. $\ln\tilde{r}_0$ (markers) calculated using Eq.~\eqref{eq:dflondon} for $\beta=0.001$ (green) and $\beta=0.005$ (red), with $d=\delta_1/\delta_2=-0.9998$ for the London model Eq.~\eqref{eq:fslondon}. The solid lines are theoretical values given by Eq.~\eqref{eq:hc1formula}, $-(\beta/4)\ln\tilde{r}_0$, for the two cases.}
\label{fig:loglondon}
\end{figure}

To better compare the single vortex in a dispersive and a flat band, we first derive $H_{c1,v}$ for the London model in the $\kappa\gg1$ limit. We expand Eq.~\eqref{eq:circular} up to the $O(A^2)$ term as an approximation of the free energy density for superconductors with dispersive bands at low temperatures:
\beq\label{eq:fslondon}
f_s(\bd{A})=-(\delta_1+\delta_2)+\frac{\delta_2}{2}\bigg(\frac{2ea}{\hbar}A\bigg)^2.
\eeq
Then the differential equation for $\tilde{A}$, Eq.~\eqref{eq:radialcos}, becomes the modified Bessel equation,
\beq\label{eq:radiallondon}
\tilde{A}''+\frac{\tilde{A}'}{\tilde{r}}-\bigg(\frac{1}{\tilde{r}^2}+1\bigg)\tilde{A}=0,
\eeq
whose (real) solution is the modified Bessel function of the second kind,
\beq\label{eq:bessel}
\tilde{A}(\tilde{r})=-\frac{\beta}{2}K_1(\tilde{r}).
\eeq
Here the prefactor is chosen to be $-\beta/2$ to satisfy Eq.~\eqref{eq:quantization} in the $\tilde{r}_0\rightarrow0$ limit, considering the asymptotic behavior of $K_1(\tilde{r})\sim \frac{1}{\tilde{r}}+\frac{\tilde{r}}{2}\ln\frac{\tilde{r}}{2}$ as $\tilde{r}\rightarrow0$. In Fig.~\ref{fig:loglondon}(a), we plot $\tilde{A}(\tilde{r})$ of Eq.~\eqref{eq:bessel} and its $B$ field, calculated from $\tilde{B}(\tilde{r})=\tilde{A}'(\tilde{r})+\tilde{A}(\tilde{r})/\tilde{r}$ for the region $\tilde{r}>\tilde{r}_0$.

To calculate the lower critical field, we equate the free energy line density of a vortex to that of the zero-field state. The free energy density difference is
\beq
\delta f^H(r)=f_n^H-f_s^H(\text{zero-field})=-\frac{1}{2}\mu_0H^2+\delta_1+\delta_2
\eeq
in the region $r<r_0$; and
\begin{align}
\delta f^H(r)=&f_s^H(r,\text{vortex})-f_s^H(r,\text{zero-field}) \nonumber\\
=&2\delta_2\bigg(\frac{ea}{\hbar}A(r)\bigg)^2+\frac{1}{2\mu_0}B(r)^2-HB(r)
\end{align}
in the region $r_0<r<R_0$. Put into dimensionless quantities, the difference of the free energy line density is
\begin{align}\label{eq:dflondon}
&\delta F^H/L=\int \derd^2r\delta f^H=\delta_2\lambda_L^2\bigg\{(1+d-2\tilde{H}^2)\pi\tilde{r}_0^2 \nonumber\\
&+\int_{\tilde{r}_0}^{\tilde{R}_0}2\pi\tilde{r}\derd \tilde{r}\big[2\tilde{A}(\tilde{r})^2+2\tilde{B}(\tilde{r})^2-4\tilde{H}\tilde{B}(\tilde{r})\big]\bigg\},
\end{align}
where $d=\delta_1/\delta_2$ and $L$ is the length along $z$ direction. Then, the critical field $\tilde{H}_{c1,v}$ is solved from $\delta F^H/L=0$.

In the literature, three approximations are made to Eq.~\eqref{eq:dflondon} to obtain an analytical expression of $H_{c1,v}$ for $\kappa\gg1$~\cite{abrikosov}. (i) First, the contribution from the normal region is neglected, since it is proportional to $\tilde{r}_0^2\sim\kappa^{-2}$. This approximation is justified as $r_0$ approaches the lattice scale, but fails if $a\ll r_0$, even if $r_0\ll \lambda_L$. For the normal core energy to be negligible, we require
\beq
(1+d-2\tilde{H}^2)\pi\tilde{r}_0^2\ll \int_{\tilde{r}_0}^{\tilde{R}_0}4\pi\tilde{r}\derd \tilde{r}\tilde{A}(\tilde{r})^2.
\eeq
Here, the $\tilde{H}^2$ term can be neglected since roughly $\tilde{H}\propto\beta$; we choose $d=-0.9998$ for the dispersive band model Eq.~\eqref{eq:fslondon}, implying that $f_s$ reaches zero at the critical vector potential $\tilde{A}_c=\sqrt{(1+d)/2}=0.01$. Then it leads to
\beq
10^{-4}\tilde{r}_0^2\ll\beta^2\ln\kappa,\,\,\,\text{or}\,\,\,\,\,\,r_0\ll100(\ln\kappa)^{1/2}a
\eeq
which holds for $r_0\gtrsim a$ only. (ii) Second, the leading contribution from $\tilde{A}^2$ and $\tilde{B}^2$ to the second line of Eq.~\eqref{eq:dflondon} is from $\tilde{r}_0<\tilde{r}<1$, so one can set $\tilde{R}_0=1$ for their integrals. In this region, $\tilde{A}(\tilde{r})^2\gg\tilde{B}(\tilde{r})^2$ (see Fig.~\ref{fig:loglondon}(a)), so the $\tilde{B}^2$ term can also be neglected. The $\tilde{H}\tilde{B}$ term can be rewritten as (the upper limit of the integral is set to $\infty$)
\beq
-4\tilde{H}\int_{\tilde{r}_0}^\infty2\pi\tilde{r}\derd \tilde{r}\tilde{B}=-4\tilde{H}\frac{\phi_0}{\lambda_L^2\hbar/ea\lambda_L}=-4\pi\beta\tilde{H}.
\eeq
As a result, the equation $\delta F^H/L=0$ reduces to
\beq\label{eq:integral}
\int_{\tilde{r}_0}^14\pi\tilde{r}\derd \tilde{r}\tilde{A}^2-4\pi\beta\tilde{H}_{c1,v}=0.
\eeq
(iii) Third, an inverse approximation, $\tilde{A}(\tilde{r})\approx-\beta/2\tilde{r}$, is made for Eq.~\eqref{eq:bessel} in the region $\tilde{r}_0<\tilde{r}<1$ (see Fig.~\ref{fig:loglondon}(a) for a comparison between the two functions). With these approximations, one then integrates Eq.~\eqref{eq:integral} to get
\beq\label{eq:hc1formula}
\tilde{H}_{c1,v}=\frac{\beta}{4}\ln\frac{\lambda_L}{r_0}.
\eeq
Putting back the units (Eq.~\eqref{eq:dimensionless}), we get
\beq
H_{c1,v}=\frac{\phi_0}{4\pi\mu_0\lambda_L^2}\ln\frac{\lambda_L}{r_0}.
\eeq

In Fig.~\ref{fig:loglondon}(b), we provide numerical calculations to support this analysis. For $r_0\gg a$, the normal core increases the free energy substantially via the first term of Eq.~\eqref{eq:dflondon}, making $\tilde{H}_{c1,v}$ larger than the value given by Eq.~\eqref{eq:hc1formula}.

\subsection{The cosine model of flat bands}
For flat bands, we solve Eq.~\eqref{eq:radialcos} for $\tilde{A}(\tilde{r})$ with boundary conditions Eq.~\eqref{eq:quantization2} and the asymptotical behavior
\beq
\tilde{A}(\tilde{r})\rightarrow -\frac{\beta}{2}K_1(\tilde{r}),\,\,\,\tilde{r}\rightarrow\infty,
\eeq
since Eq.~\eqref{eq:radialcos} asymptotically becomes Eq.~\eqref{eq:radiallondon} at large $\tilde{r}$. We focus on the case of a small core size but larger than the lattice constant, $r_0\gtrsim a$, since flat-band superconductors typically have a small coherence length. We exclude the case $r_0<a$ because its physical meaning is unclear.

\begin{figure}[t!]
\centering
\includegraphics[width=0.4\textwidth]{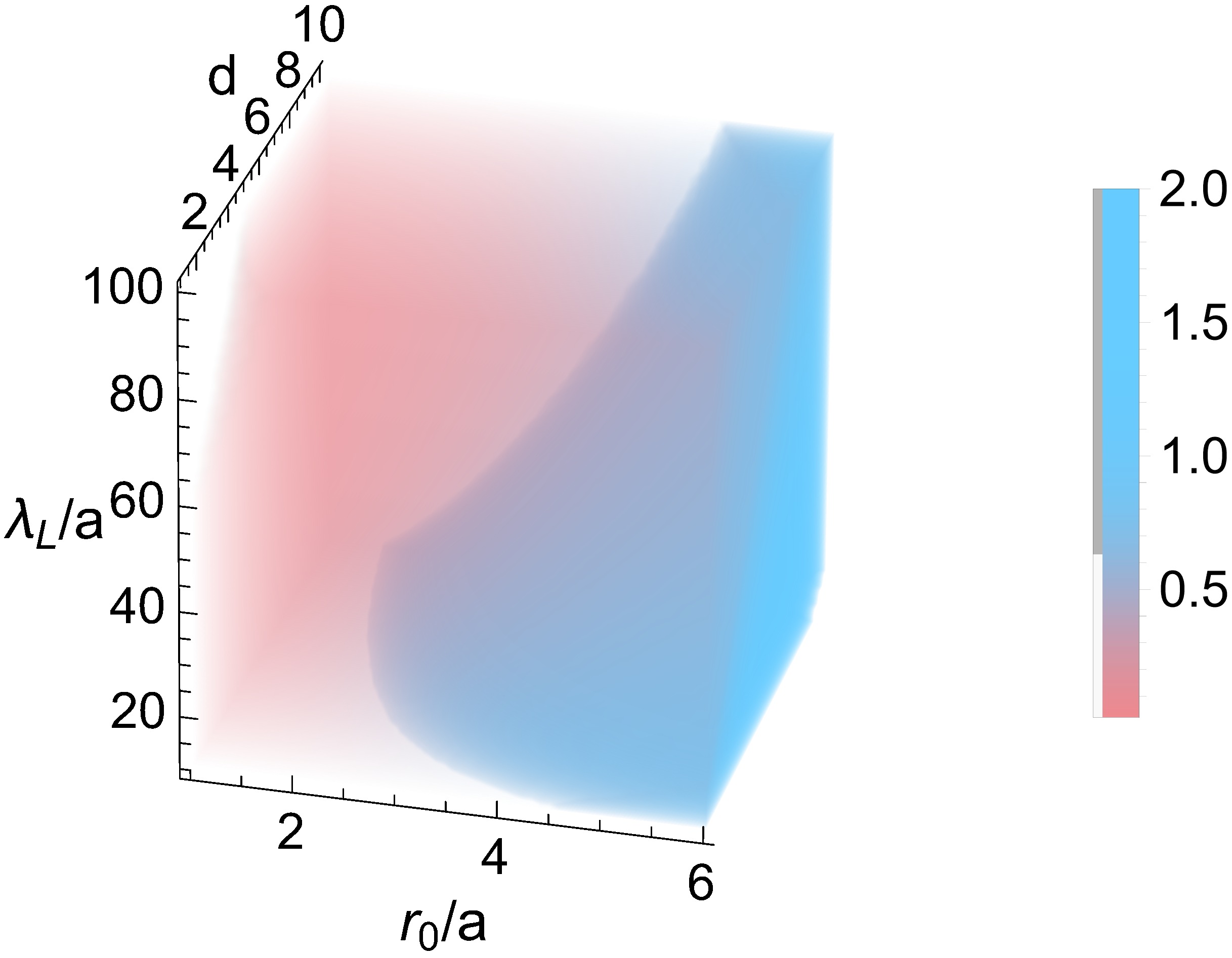}
\caption{Density plot of $\tilde{H}_{c1,v}$, Eq.~\eqref{eq:hc1vcos}, as a function of $r_0,d,\lambda_L$, with high (low) opacity on the $\tilde{H}_{c1,v}>\tilde{H}_{c1,w}$ ($\tilde{H}_{c1,v}<\tilde{H}_{c1,w}$) side ($\tilde{H}_{c1,w}\approx 0.63$).}
\label{fig:cube}
\end{figure}

We notice that the two equations, Eq.~\eqref{eq:radialcos} and \eqref{eq:radiallondon} are only substantially different as $\tilde{A}$ reaches the order of $~0.1$, which is exactly when $\tilde{r}\sim a/\lambda_L$. Then, since we consider $\tilde{r}_0>a/\lambda_L$ only, the solutions of $\tilde{A}(\tilde{r})$ to the two equations must slightly differ near the core only (this slight difference can be checked numerically, e.g., using the finite-difference method to solve Eq.~\eqref{eq:radialcos}). Therefore, in the entire domain of interest, $\tilde{r}_0<\tilde{r}<\infty$, we use the Bessel solution, Eq.~\eqref{eq:bessel}, to approximate that to Eq.~\eqref{eq:radialcos}.

The free energy line density difference for flat bands is
\begin{align}\label{eq:dfcosine}
&\delta F^H/L=\int \derd^2r\delta f^H=\delta_2\lambda_L^2\bigg\{(1+d-2\tilde{H}^2)\pi\tilde{r}_0^2 \\
&+\int_{\tilde{r}_0}^{\tilde{R}_0}2\pi\tilde{r}\derd \tilde{r}\big[1-\cos(2\tilde{A}(\tilde{r}))+2\tilde{B}(\tilde{r})^2-4\tilde{H}\tilde{B}(\tilde{r})\big]\bigg\}. \nonumber
\end{align}
Comparing with the dispersive-band case, Eq.~\eqref{eq:dflondon}, approximations (ii) and (iii) are still valid, whereas (i) fails since $d>1$ for flat bands at low temperatures. This leads to a large normal core energy, making Eq.~\eqref{eq:hc1formula} no longer hold ($\tilde{H}_{c1,v}$ of flat bands is much larger than it). Then, using the same method as described for the London model, an analytical expression of $\tilde{H}_{c1,v}$ can be obtained by solving $\delta F^H/L=0$ as a quadratic equation for $\tilde{H}_{c1,v}$. We obtain
\beq\label{eq:hc1vcos}
\tilde{H}_{c1,v}=-\frac{\beta}{\tilde{r}_0^2}+\sqrt{\frac{\beta^2}{\tilde{r}_0^4}+\frac{1+d}{2}-\frac{\beta^2}{2\tilde{r}_0^2}\ln\tilde{r}_0}.
\eeq
In Fig.~\ref{fig:cube}, we plot it as a function of $r_0,d,\lambda_L$ for small values of $1\leq r_0/a\leq 6$ and $1\leq d\leq 10$. Setting $\tilde{H}_{c1,v}(r_0,d,\lambda_L)$ equal to a constant, e.g., $\tilde{H}_{c1,w}\approx 0.63$ of the wall phase, yields a surface in the 3-dimensional parameter space (see the boundary surface between two regions with opacity contrast in Fig.~\ref{fig:cube}). Two sections of this surface, $\lambda_L=20a$ and $100a$, are shown in the main text as Fig.~\ref{fig:diagram}.

For small $\tilde{H}_{c1,v}$ values, we can also neglect the $\tilde{H}^2$ term in Eq.~\eqref{eq:dfcosine} and obtain a much simpler formula for $\tilde{H}_{c1,v}$:
\beq
\tilde{H}_{c1,v}\approx\frac{\beta}{4}\left[(1+d)\left(\frac{r_0}{a}\right)^2+\ln\frac{\lambda_L}{r_0}\right].
\eeq
Therefore, $\tilde{H}_{c1,v}$ approximately scales with $\beta$ up to a logarithmic correction. With the units put back (Eq.~\eqref{eq:dimensionless}), this implies that approximately $H_{c1,v}\propto 1/\lambda_L^2$.
%


%

\end{document}